\documentclass[sigconf]{aamas} 

\usepackage{balance} 
\usepackage{packages}
\usepackage{commands}
\usepackage{mymacros}

\setcopyright{ifaamas}
\acmConference[AAMAS '22]{Proc.\@ of the 21st International Conference
on Autonomous Agents and Multiagent Systems (AAMAS 2022)}{May 9--13, 2022}
{Online}{P.~Faliszewski, V.~Mascardi, C.~Pelachaud,
M.E.~Taylor (eds.)}
\copyrightyear{2022}
\acmYear{2022}
\acmDOI{}
\acmPrice{}
\acmISBN{}

\acmSubmissionID{776}

\title{Robust No-Regret Learning in Min-Max Stackelberg Games}

\author{Denizalp Goktas}
\affiliation{
  \institution{Brown University}
  \department{Computer Science}
  \city{Providence}
  \state{Rhode Island}
  \country{USA}}
\email{denizalp_goktas@brown.edu}

\author{Jiayi Zhao}
\affiliation{
  \institution{Pomona College}
  \department{Computer Science}
  \city{Claremont}
  \state{CA}
  \country{USA}}
\email{jzae2019@mymail.pomona.edu}

\author{Amy Greenwald}
\affiliation{
  \institution{Brown University}
  \department{Computer Science}
  \city{Providence}
  \state{Rhode Island}
  \country{USA}}
\email{amy_greenwald@brown.edu}

\begin{abstract}
    The behavior of no-regret learning algorithms is well understood in two-player min-max (i.e, zero-sum) games. In this paper, we investigate the behavior of no-regret learning in min-max games \emph{with dependent strategy sets}, where the strategy of the first player constrains the behavior of the second. Such games are best understood as sequential, i.e., min-max Stackelberg, games. We consider two settings, one in which only the first player chooses their actions using a no-regret algorithm while the second player best responds, and one in which both players use no-regret algorithms. For the former case, we show that no-regret dynamics converge to a Stackelberg equilibrium. For the latter case, we introduce a new type of regret, which we call Lagrangian regret, and show that if both players minimize their Lagrangian regrets, then play converges to a Stackelberg equilibrium. We then observe that online mirror descent (OMD) dynamics in these two settings correspond respectively to a known nested (i.e., sequential) gradient descent-ascent (GDA) algorithm and a new simultaneous GDA-like algorithm, thereby establishing convergence of these algorithms to Stackelberg equilibrium. Finally, we analyze the robustness of OMD dynamics to perturbations by investigating online min-max Stackelberg games. We prove that OMD dynamics are robust for a large class of online min-max games with independent strategy sets. In the dependent case, we demonstrate the robustness of OMD dynamics experimentally by simulating them in online Fisher markets, a canonical example of a min-max Stackelberg game with dependent strategy sets.
\end{abstract}



\newcommand{\BibTeX}{\rm B\kern-.05em{\sc i\kern-.025em b}\kern-.08em\TeX}

\begin{document}


\pagestyle{fancy}
\fancyhead{}


\maketitle 

\section{Introduction}
\label{sec:intro}

Min-max optimization problems (i.e., zero-sum games) have been attracting a great deal of attention recently because of their applicability to problems in fairness in machine learning \cite{dai2019kernel, edwards2016censoring, madras2018learning, sattigeri2018fairness}, generative adversarial imitation learning \cite{cai2019global, hamedani2018iteration}, reinforcement learning \cite{dai2018rl}, generative adversarial learning \cite{sanjabi2018convergence}, \amy{you should cite Goodfellow here. wasn't it his idea originally?} adversarial learning \cite{sinha2020certifying}, and statistical learning, e.g., learning parameters of exponential families \cite{dai2019kernel}. 
These problems are often modelled as \mydef{min-max games}, i.e., constrained min-max optimization problems of the form:
$\min_{\outer \in \outerset} \max_{\inner \in \innerset} \obj(\outer, \inner)$,
where $\obj: \outerset \times \innerset \to \R$ is continuous, and $\outerset \subset \R^\outerdim$ and $\innerset \subset \R^\innerdim$ are non-empty and compact.
In \mydef{convex-concave min-max games}, where $\obj$ is convex in $\outer$ and concave in $\inner$, von Neumann and Morgenstern's seminal minimax theorem holds  \cite{neumann1928theorie}: i.e.,
$\min_{\outer \in \outerset} \max_{\inner \in \innerset} \obj(\outer, \inner) = \max_{\inner \in \innerset} \min_{\outer \in \outerset} \obj(\outer, \inner)$, guaranteeing the existence of a saddle point, i.e., a point that is simultaneously a minimum of $\obj$ in the $\outer$-direction and a maximum of $\obj$ in the $\inner$-direction.
Because of the minimax theorem, we can interpret the constrained optimization problem as a simultaneous-move, zero-sum game, where $\inner^*$ (resp. $\outer^*$) is a best-response of the outer (resp. inner) player to the other's action $\outer^*$ (resp. $\inner^*)$, in which case a saddle point is also called a minimax point or a Nash equilibrium.

In this paper, we study 
\mydef{min-max Stackelberg games} \cite{goktas2021minmax}, i.e., constrained min-max optimization problems \emph{with dependent feasible sets\/} of the form: $\min_{\outer \in \outerset} \max_{\inner \in \innerset : \constr(\outer, \inner) \geq \zeros} \obj(\outer, \inner)$,
where $\obj: \outerset \times \innerset \to \R$ is continuous, $\outerset \subset \R^\outerdim$ and $\innerset \subset \R^\innerdim$ are non-empty and compact, and $\constr(\outer, \inner) = \left(\constr[1](\outer, \inner), \hdots, \constr[\numconstrs](\outer, \inner) \right)^T$ with $\constr[\numconstr]: \outerset \times \innerset \to \R$.
\citeauthor{goktas2021minmax} observe that the minimax theorem does not hold in these games \cite{goktas2021minmax}.
As a result, such games are more appropriately viewed as sequential, i.e., Stackelberg, games for which the relevant solution concept is the Stackelberg equilibrium,%
\footnote{Alternatively, one could view such games as pseudo-games (also known as abstract economies) \cite{arrow-debreu}, in which players move simultaneously under the unreasonable assumption that the moves they make will satisfy the game's dependency constraints.
Under this view, the relevant solution concept is generalized Nash equilibrium \cite{facchinei2007generalized, facchinei2010generalized}.}
where the outer player chooses $\hat{\outer} \in \outerset$ before the inner player responds with their choice of $\inner(\hat{\outer}) \in \innerset$ s.t.\ $\constr(\hat{\outer}, \inner(\hat{\outer})) \geq \zeros$.
The outer player's objective, which is referred to as their \mydef{value function} in the economics literature \cite{milgrom2002envelope} and which they seek to minimize, is defined as $\val[\outerset](\outer) = \max_{\inner \in \innerset : \constr(\outer, \inner) \geq \zeros} \obj(\outer, \inner)$.
The inner player's value function, $\val[\innerset]: \outerset \to \R$, which they seek to maximize, is simply the objective function of the game, given the outer player's action $\hat{\outer}$: i.e., $\val[\innerset](\inner; \hat{\outer}) = \obj(\hat{\outer}, \inner)$.

\citeauthor{goktas2021minmax} \cite{goktas2021minmax} proposed a polynomial-time first-order method by which to compute Stackelberg equilibria, which they called \mydef{nested gradient descent ascent (GDA)}.
This method can be understood as an algorithm a third party might run to find an equilibrium, or as a game dynamic that the players might employ if their long-run goal were to reach an equilibrium.
Rather than assume that players are jointly working towards the goal of reaching an equilibrium, it is often more reasonable to assume that they play so as to not regret their decisions: i.e., that they employ a \mydef{no-regret learning algorithm}, which minimizes their loss in hindsight.
It is well known that when both players in a repeated min-max game are no-regret learners, the players' strategy profile over time converges to a Nash equilibrium in average iterates: i.e., 
empirical play converges to a Nash equilibrium (e.g., \cite{freund1996game}).

In this paper, we investigate no-regret learning dynamics in repeated min-max Stackelberg games.
We assume both an asymmetric and a symmetric setting.
In the asymmetric setting, the outer player is a no-regret learner while the inner player best responds; in the symmetric setting, both players are no-regret learners.
In the asymmetric case, we show that if the outer player uses a no-regret algorithm that achieves $\varepsilon$-{asymmetric} regret, then the outer player's empirical play converges to their $\varepsilon$-Stackelberg equilibrium strategy.
In the symmetric case, we introduce a new type of regret, which we call Lagrangian regret,%
\footnote{We note that similar notions of Lagrangian regret have been used in other online learning settings (e.g., \cite{bechavod2020metric}), but to our knowledge, ours is the first game-theoretic analysis of Lagrangian regret minimization.}
which assumes access to a solution oracle for the optimal KKT multipliers of the game's constraints.
We then show that if both players use no-regret algorithms that achieve $\varepsilon$-Lagrangian regrets, then the players' empirical play converges to an $\varepsilon$-Stackelberg equilibrium.

Next, we restrict our attention to a specific no-regret dynamic, namely online mirror descent (OMD)~\cite{nemirovski2004prox}.
Doing so yields two algorithms, max-oracle mirror descent (max-oracle MD) and nested mirror descent ascent (nested MDA) in the asymmetric setting, and a new simultaneous GDA-like algorithm \cite{nedic2009gda} in the symmetric setting, which we call Lagrangian mirror descent ascent (LMDA).
The first two algorithms converge to $\varepsilon$-Stackelberg equilibrium in $O(\nicefrac{1}{\varepsilon^2})$ and $O(\nicefrac{1}{\varepsilon^3})$ iterations, respectively, and the third, in $O(\nicefrac{1}{\varepsilon^2})$, when a Lagrangian solution oracle exists.
As max-oracle gradient~\cite{goktas2021minmax,jin2020local} and nested GDA~\cite{goktas2021minmax} are special cases of max-oracle MD and nested MDA, respectively, our convergence bounds complement \citeauthor{goktas2021minmax}'s best iterate convergence results, now proving average iterate convergence for both algorithms.
Furthermore, our result on LMDA's convergence rate suggests the computational superiority of LMDA over nested GDA, when a Lagrangian solution oracle exists.
We also note that even when such an oracle does not exist, the Lagrangian solution can be treated as a hyperparameter of the algorithm allowing for a significant speed up in computation.

Finally, we analyze the robustness of OMD dynamics 
by investigating online min-max Stackelberg games{, i.e., min-max Stackelberg games with arbitrary objective and constraint functions from one time step to the next}.
We prove that OMD dynamics are robust, in that even when the game changes, OMD dynamics track the changing equilibria closely, in a large class of online min-max games with independent strategy sets.
In the dependent strategy set case, we demonstrate the robustness of OMD dynamics experimentally by simulating online Fisher markets, a canonical example of an (online) min-max Stackelberg game (with dependent strategy sets) \cite{goktas2021minmax}.
Even when the Fisher market changes every time step, our OMD dynamics track the changing equilibria closely.
These results are somewhat surprising, because optimization problems can be highly sensitive to perturbations of their inputs \cite{ben2000robust}.

Our findings can be summarized as follows:
\begin{itemize}[topsep=0pt]
    
    \item In repeated min-max Stackelberg games, when the outer player is a no-regret learner and the inner-player best-responds, the average of the outer player's strategies converges to their Stackelberg equilibrium strategy.
    
    \item We introduce a new type of regret we call Lagrangian regret and show that in repeated min-max Stackelberg games when both players minimize Lagrangian regret, the average of the players' strategies converge to a Stackelberg equilibrium.
    
    \item We provide convergence guarantees for max-oracle MD and nested MDA to an $\varepsilon$-Stackelberg equilibrium in $O(\nicefrac{1}{\varepsilon^2})$ and $O(\nicefrac{1}{\varepsilon^3})$ in average iterates, respectively.
    
    \item We introduce a simultaneous GDA-like algorithm, which we call LMDA, and prove that its average iterates converge to an $\varepsilon$-Stackelberg equilibrium in $O(\nicefrac{1}{\varepsilon^2})$ iterations.
    
    \item We prove that max-oracle MD and LMDA are robust to perturbations in a large class of online min-max games (with independent strategy sets).
    
    \item We run experiments with Fisher markets which suggest that max-oracle MD and LMDA are robust to perturbations in 
    these online min-max Stackelberg games.
\end{itemize}


\paragraph{Related Work}




Stackelberg games \cite{stackelberg1934marktform} have found important applications in the domain of security (e.g., \cite{nguyen2016ssg, sinha2018stackelberg}) and environmental protection (e.g., \cite{fang2016sgsg}).
These applications have thus far been modelled as Stackelberg games with independent strategy sets.
Yet, the increased expressiveness of Stackelberg games with dependent strategy sets may make them a better model of the real world, as they provide the leader with more power to achieve a better outcome by constraining the follower's choices.

The study of algorithms that compute competitive equilibria in Fisher markets was initiated by \citeauthor{devanur2002market} \cite{devanur2002market}, who provided a polynomial-time method for solving these markets assuming linear utilities.
More recently, \citeauthor{cheung2019tracing} \cite{cheung2019tracing} studied two price adjustment processes, t\^atonnement and proportional response dynamics, in dynamic Fisher markets and showed that these price adjustment processes track the equilibrium of Fisher markets closely even when the market is subject to change.

The computation and learning of Nash and generalized Nash equilibrium in min-max games (with independent strategy sets) has been attracting a great deal of attention recently, because of the relevance of these problems to machine learning \cite{daskalakis2020complexity, abernethy2019last, daskalakis2018last, daskalakis2018limit, diakonikolas2021efficient}, specifically generative adversarial learning \cite{goodfellow2014generative}. \amy{add references to GANs here, including Ian Goodfellow}

\section{Mathematical Preliminaries}
\label{sec:prelim}

\paragraph{Notation}

We use Roman uppercase letters to denote sets (e.g., $X$),
bold uppercase letters to denote matrices (e.g., $\allocation$), bold lowercase letters to denote vectors (e.g., $\price$), and Roman lowercase letters to denote scalar quantities, (e.g., $c$).
We denote the $i$th row vector of a matrix (e.g., $\allocation$) by the corresponding bold lowercase letter with subscript $i$ (e.g., $\allocation[\buyer])$.
Similarly, we denote the $j$th entry of a vector (e.g., $\price$ or $\allocation[\buyer]$) by the corresponding Roman lowercase letter with subscript $j$ (e.g., $\price[\good]$ or $\allocation[\buyer][\good]$).
We denote the vector of ones of size $\numbuyers$ by $\ones[\numbuyers]$.
We denote the set of integers $\left\{1, \hdots, n\right\}$ by $[n]$, the set of natural numbers by $\N$, the set of positive natural numbers by $\N_+$ the set of real numbers by $\R$, the set of non-negative real numbers by $\R_+$, and the set of strictly positive real numbers by $\R_{++}$.
We denote the orthogonal projection operator onto a convex set $C$ by $\project[C]$, i.e., $\project[C](\x) = \argmin_{\y \in C} \left\|\x - \y \right\|^2$. 
Given a sequence of iterates $\{ \z^{(\iter)} \}_{\iter =1}^\numiters \subset Z$, we denote the average iterate $\bar{\z}^{(\numiters)} = \frac{1}{\numiters} \sum_{\iter =1 }^\numiters \z^{(\iter)}$.

\paragraph{Game Definitions}

A \mydef{min-max Stackelberg game}, $(\outerset, \innerset, \obj, \constr)$, is a two-player, zero-sum game, where one player, who we call the \mydef{outer}
player (resp.\ the \mydef{inner}
player), is trying to minimize their loss (resp.\ maximize their gain), defined by a continuous \mydef{objective function} $\obj: X \times Y \rightarrow \mathbb{R}$, by choosing a strategy from their non-empty and compact \mydef{strategy set} $\outerset \subset \R^\outerdim$, and (resp. $\innerset \subset \R^\innerdim$) s.t.\ $\constr(\outer, \inner) \geq 0$ where $\constr(\outer, \inner) = \left(\constr[1](\outer, \inner), \hdots, \constr[\numconstrs](\outer, \inner) \right)^T$ with $\constr[\numconstr]: \outerset \times \innerset \to \R$ continuous.
A strategy profile $(\outer, \inner) \in \outerset \times \innerset$ is said to be \mydef{feasible} iff for all $\numconstr \in [\numconstrs]$, $\constr[\numconstr](\outer, \inner) \geq 0$.
The function $\obj$ maps a pair of strategies taken by the players $(\outer, \inner) \in \outerset \times \innerset$ to a real value (i.e., a payoff), which represents the loss (resp.\ the gain) of the outer player (resp.\ the inner player).
A min-max game is said to be convex-concave if the objective function $\obj$ is convex-concave and $\outerset$ and $\innerset$ are convex sets.

The relevant solution concept for Stackelberg games is the \mydef{Stackelberg equilibrium (SE)}:
A strategy profile $\left( \outer^{*}, \inner^{*} \right) \in \outerset \times \innerset$ s.t.\ $\constr \left( \outer^{*}, \inner^{*} \right) \geq \zeros$ is an $(\epsilon, \delta)$-SE if
$\max_{\inner \in \innerset : \constr \left( \outer^{*}, \inner \right) \geq 0} \obj \left( \outer^{*}, \inner \right) - \delta \leq \obj \left( \outer^{*}, \inner^{*} \right) \leq \min_{\outer \in \outerset} \max_{\inner \in \innerset : \constr(\outer, \inner) \geq 0} \obj \left( \outer, \inner \right) + \epsilon$.
%
Intuitively, a $(\varepsilon, \delta)$-SE is a point at which the outer player's (resp.\ inner player's) payoff is no more than $\varepsilon$ (resp.\ $\delta$) away from its optimum.
A $(0,0)$-SE is guaranteed to exist in min-max Stackelberg games \cite{goktas2021minmax}.
Note that when $\constr(\outer, \inner) \geq \zeros$, for all $(\outer, \inner) \in \outerset \times \innerset$, the game reduces to a min-max game (with independent strategy sets).

In a min-max Stackelberg game, the outer player's \mydef{best-response set} $\br[\outerset] \subset \outerset$, defined as $\br[\outerset] = \argmin_{\outer \in \outerset} \val[\outerset](\outer)$, is independent of the inner player's strategy, while the inner player's \mydef{best-response correspondence} $\br[\innerset] : \outerset \rightrightarrows \innerset$, defined as $\br[\innerset](\outer) = \argmax_{\inner \in \innerset: \constr(\outer, \inner) \geq 0} \val[\innerset](\inner; \outer)$,
depends on the outer player's strategy.
A $(0,0)$-Stackelberg equilibrium $(\outer^*, \inner^*) \in \outerset \times \innerset$ is then a tuple of strategies such that $(\outer^*, \inner^*) \in \br[\outerset] \times \br[\innerset](\outer^*)$.

An \mydef{online min-max Stackelberg game}, $\left\{ \left( \outerset, \innerset, \obj[\iter], \constr[][\iter] \right) \right\}$, 
is a sequence of min-max Stackelberg games played for $\numiters$ time periods.
We define the players' value functions at time $\iter$ in a online min-max Stackelberg game in terms of $\obj[\iter]$ and $\constr[][\iter]$.
Note that when $\constr[][\iter](\outer, \inner) \geq 0$ for all $\outer \in \outerset, \inner \in \innerset$ and all time periods $\iter \in \iters$, the game reduces to a online min-max game (with independent strategy sets).
Moreover, if for all $\iter, \iter' \in \iters, \obj[\iter] = \obj[\iter']$, and $\constr[][\iter] = \constr[][\iter']$, then the game reduces to a \mydef{repeated min-max Stackelberg game}, which we denote simply by $(\outerset, \innerset, \obj, \constr)$.

\paragraph{Assumptions}
All the theoretical results on min-max Stackelberg games in this paper rely on the following assumption(s):
\sdeni{
}{

\begin{assumption}
\label{main-assum}
1.~(Slater's condition)
$\forall \outer \in \outerset, \exists \widehat{\inner} \in \innerset$ s.t.\ $g_{\numconstr}(\outer, \widehat{\inner}) > 0$, for all $\numconstr \in [\numconstrs]$;
2.~$\grad[\outer] f, \grad[\outer] \constr[1], \ldots, \grad[\outer] \constr[\numconstrs]$ are continuous;
and 3.a.~$\obj$ is continuous and convex-concave, 3.b.~$\mu \constr[1](\outer, \inner), \ldots,$ $\mu \constr[\numconstrs](\outer, \inner)$ are continuous, convex in $(\mu, \outer)$ over the set $\R_+ \times \outerset$, for all $\inner \in \innerset$, and concave in $\inner$ over the set $\innerset$, for all $(\mu, \outer) \in \R_+ \times \outerset$.
\end{assumption}
}

We note that these assumptions are in line with previous work geared towards solving min-max Stackelberg games 
\cite{goktas2021minmax}.
Part 1 of \Cref{main-assum},
Slater's condition, is a standard constraint qualification condition \cite{boyd2004convex}, which is needed to derive the optimality conditions for the inner player's maximization problem; without it the problem becomes analytically intractable.
Part 2 of \Cref{main-assum} ensures that the value function of the outer player is continuous and convex (\cite{goktas2021minmax}, Proposition A1), so that the problem affords an efficient solution.
Part 3 of \Cref{main-assum} can be replaced by a weaker, subgradient boundedness assumption; however, for simplicity, we assume this stronger condition.
Finally, Part 4 of \Cref{main-assum} guarantees that projections are polynomial-time operations.

Under \Cref{main-assum}, the following property holds of the outer player's value function.

\begin{proposition}[\cite{goktas2021minmax}, Proposition B.1]
\label{thm:convex-value-func}
Consider a min-max Stackelberg game $(\outerset, \innerset, \obj, \constr)$ and suppose that \Cref{main-assum} holds, then the outer player's value function $\val(\outer) = \max_{\inner \in \innerset : \constr(\outer, \inner) \geq \zeros} \obj(\outer, \inner)$ is continuous and convex.
\end{proposition}

\paragraph{Additional Definitions}
%
Given two normed spaces $(\outerset, \|\cdot \|)$ and $(\innerset, \|\cdot \|)$, the function $\obj: \outerset \to \innerset$ is
$\lipschitz[\obj]$-\mydef{Lipschitz-continuous} iff $\forall \outer_1, \outer_2 \in X, \left\| \obj(\outer_1) - \obj(\outer_2) \right\| \leq \lipschitz[\obj] \left\| \outer_1 - \outer_2 \right\|$.
If the gradient of $\obj$, $\grad \obj$, is $\lipschitz[\grad \obj]$-Lipschitz-continuous, we refer to $\obj$ as $\lipschitz[\grad \obj]$-\mydef{Lipschitz-smooth}.
A function $\obj: A \to \R$ is $\mu$-\mydef{strongly convex} if $\obj(\outer_1) \geq \obj(\outer_2) + \left< \grad[\outer] \obj(\outer_2), \outer_1 - \outer_2 \right> + \nicefrac{\mu}{2} \left\| \outer_1 - \outer_1 \right\|^2$, and $\mu$-\mydef{strongly concave} if $-\obj$ is $\mu$-strongly convex.

\paragraph{Online Convex Optimization}


An \mydef{online convex optimization problem (OCP)} is a decision problem in a dynamic environment which comprises a finite time horizon $\numiters$, a compact, convex feasible set $\outerset$, and a sequence of convex differentiable loss functions $\{\loss[][\iter] \}_{\iter = 1}^\numiters$, where $\loss[][\iter]: \outerset \to \R$ for all $\iter \in [\numiters]$.
A solution to an OCP is a sequence $\{ \outer^{(\iter)} \}_{\iter = 1}^\numiters$ with each $\outer^{(\iter)} \in \outerset$.
A preferred solution is one that minimizes \mydef{average regret}, given by
%
%
$\regret[][\numiters](\left\{ \outer^{\iter} \right\}, \outer) = \sum_{\iter = 1}^\numiters \frac{1}{\numiters}\loss[][\iter](\outer^{(\iter)}) - \sum_{\iter = 1}^\numiters \frac{1}{\numiters} \loss[][\iter](\outer)$,
for all $\outer \in \outerset$.
Overloading notation, we also write $\regret[][\numiters](\left\{ \outer^{\iter} \right\}) = \max_{\outer \in \outerset} \regret[][\numiters](\left\{ \outer^{\iter} \right\}, \outer)$.
An algorithm $\algo$ 
that takes as input a sequence of loss functions and outputs decisions such that  $\regret[][\numiters](\algo(\{\loss[][\iter] \}) \to 0$ 
as $\numiters \to \infty$ is called a \mydef{no-regret algorithm}.

For any differentiable convex function $\regul: \outerset \to \R$, the \mydef{Bregman divergence} between two vectors $\w, \u \in \outerset$ is defined as follows:
    $\bregman[\regul](\w||\u)=\regul(\w)-(\regul(\u)+\left<\grad \regul(\u), (\w-\u)\right>$.
One first-order no-regret learning algorithm is \mydef{Online Mirror Descent (OMD)}, defined as follows for some initial iterate $\outer^{(0)} \in \outerset$, a fixed learning rate $\learnrate[ ] > 0$, and a strongly convex regularizer $\regul$:
$\outer^{(\iter+1)} = \argmin_{\outer \in \outerset} \left< \grad[\outer] \loss[][\iter](\outer^{(\iter)}), \outer \right> + \frac{1}{2\learnrate[ ]} \bregman[\regul](\outer || \outer^{(\iter)})$.
%
When $\regul(\outer) = \frac{1}{2} \left\|\outer \right\|^2_2$, OMD reduces to \mydef{projected online gradient descent (OGD)}, given by the update rule:
    $\outer^{(\iter + 1)} = \proj[\outerset] \left(\outer^{(\iter)} - \eta \grad[\outer] \loss[ ][\iter] (\outer^{(\iter)}) \right)$. 
%
The next theorem bounds the \mydef{average regret} of OMD \cite{kakade2012regularization}:


\begin{theorem}
Suppose that the OMD algorithm generates a sequence of iterates $\{ \outer^{(\iter)}\}$ when run with a $1$-strongly convex regularizer $\regul$%
\footnote{This assumption is without loss of generality, since any $m$-strongly-convex regularizer can be transformed into a $1$-strongly-convex regularizer}.
Let $c = \max_{\outer \in \outerset, \iter \in \iters} \bregman[\regul](\outer || \outer^{(\iter)})$, and let $\{\loss[ ][\iter] \}$ be a sequence of functions s.t.\ for all $\iter \in \N_+$, $\loss[ ][\iter]: \R^\outerdim \to \R$ is $\lipschitz$-Lipschitz w.r.t. the dual norm $\left\| \cdot \right\|_*$.
Then, if $\learnrate[ ] = \frac{c}{\lipschitz\sqrt{2T}}$, OMD achieves average regret bounded as follows:
$\regret[][\numiters](\left\{ \outer^{\iter} \right\}) \leq c  \lipschitz \sqrt{\nicefrac{2}{\numiters}}$.
\end{theorem}

\section{No-Regret Learning Dynamics}
\label{sec:no-regret}

In Stackelberg games, the outer player chooses their strategy assuming the inner player will best respond.
When both players' choices are optimal, the outcome is a Stackelberg equilibrium. 

In this section, we study no-regret learning dynamics in repeated min-max Stackelberg games in two settings: an \mydef{asymmetric} one in which the outer player is a no-regret learner while the inner player best-responds, and a \mydef{symmetric} one in which both players are no-regret learners.
Our main results are: 1.~In the asymmetric setting, if the outer player employs an asymmetric-regret-minimizing algorithm, play converges to a Stackelberg equilibrium, and 2.~in the symmetric setting, if both players employ a no-Lagrangian-regret algorithm, play converges to a Stackelberg equilibrium.

\subsection{Asymmetric Learning Setting}


We first consider an asymmetric setting in which the inner player best responds to the strategy picked by the outer player, while the outer player employs a no-regret learning algorithm. 
In min-max Stackelberg games, the two players are adversaries, so this best-response assumption corresponds to the worst case.
In many real-world applications,  we seek optimal strategies for the outer player, e.g., in security games we are interested in an optimal strategy for the defender/outer player, not the attacker/inner player~\cite{kar2017trends}.
Assuming a strong inner player allows us to learn more robust 
strategies for the outer player.

Given $\outer \in \outerset$, let $\inner^*(\outer) \in \br[\innerset](\outer)$,
and consider an online min-max Stackelberg game $\left\{\left( \outerset, \innerset, \obj[\iter], \constr[][\iter] \right) \right\}$.
In an asymmetric setting, the outer player's regret is the difference between the cumulative loss of their sequence of strategies $\{\outer[][\iter]\}$ (to which the inner player best responds), and the smallest cumulative loss that the outer player could have achieved by playing a fixed strategy $\outer \in \outerset$ (again, to which the inner player best responds), i.e., $\frac{1}{\numiters}\sum_{\iter = 1}^\numiters \obj[\iter](\outer[][\iter], \inner^*(\outer[][\iter])) - \sum_{\iter =1}^\numiters \frac{1}{\numiters} \obj[\iter](\outer, \inner^*(\outer))$. We call this regret the \mydef{asymmetric regret},
and express it in terms of the outer player's value function $\val[\outerset]$:
%
    $\pesregret[\outerset][\numiters] \left( \left\{ \outer[][\iter] \right\}, \outer \right) = \frac{1}{\numiters}\sum_{\iter = 1}^\numiters \val[\outerset][\iter](\outer[][\iter]) - \sum_{\iter =1}^\numiters \frac{1}{\numiters} \val[\outerset][\iter](\outer)$.
%
%
As above, we overload notation and write \\ $\pesregret[\outerset][\numiters] \left( \left\{ \outer[][\iter] \right\} \right) = \max_{\outer \in \outerset} \pesregret[\outerset][\numiters] \left( \left\{ \outer[][\iter] \right\}, \outer \right)$.

The main theorem%
\footnote{The proofs of all mathematical claims in this section can be found in \Cref{sec_app:proofs}.}
in this section states the following: assuming the inner player best responds to the strategies of the outer player, if the outer player employs a no-regret algorithm, then the outer player's average strategy converges to their part of a Stackelberg equilibrium strategy.

\begin{theorem}
\label{thm:pes-regret-bound}
Consider a repeated min-max Stackelberg game $(\outerset, \innerset, \obj, \constr)$, and suppose the outer player plays a sequence of strategies $\{\outer[][\iter]\}$. 
If, after $\numiters$ iterations, the outer player's asymmetric regret is bounded by $\varepsilon$, i.e.,
$\pesregret[\outerset][\numiters] \left( \left\{ \outer[][\iter] \right\} \right) \le \epsilon$,
then $\left( \avgouter[][\numiters],  \inner^*(\avgouter[][\numiters]) \right)$ is a $(\varepsilon, 0)$-Stackelberg equilibrium, where $\inner^*(\avgouter[][\numiters]) \in \br[\innerset](\avgouter[][\numiters])$.
\end{theorem}

We remark that although the definition of asymmetric regret looks similar to the standard definition of regret, its structure is very different.
\Cref{thm:convex-value-func} is required to ensure that the time-averaged value function $\sum_{\iter =1}^\numiters \val[][\iter](\outer)$ is convex in $\outer$.


\subsection{Symmetric Learning Setting} 

We now turn our attention to a setting in which both players are no-regret learners.
The most straightforward way to define regret is by considering the outer and inner players' ``vanilla'' regrets, respectively: 
$\regret[\outerset][\numiters] \left( \{\outer[][\iter]\}, \outer \right) = \frac{1}{\numiters}\sum_{\iter = 1}^\numiters \obj[\iter](\outer[][\iter], \inner[][\iter]) - \frac{1}{\numiters} \sum_{\iter =1}^\numiters  \obj[\iter](\outer, \inner[][\iter])$ and $\regret[\innerset][\numiters] \left( \{\inner[][\iter]\}, \inner \right) = \frac{1}{\numiters} \sum_{\iter =1}^\numiters \obj[\iter](\outer[][\iter], \inner) -  \frac{1}{\numiters}\sum_{\iter = 1}^\numiters \obj[\iter](\outer[][\iter], \inner[][\iter]) $.
In convex-concave min-max games (with independent strategy sets), when both players minimize these regrets, 
the players' average strategies converge to Nash equilibrium.
In min-max Stackelberg games (with dependent strategy sets), however,
convergence to a Stackelberg equilibrium is not guaranteed.

\begin{example}
Consider the min-max Stackelberg game $\min_{\outer[ ] \in [-1, 1]} \\ \max_{\inner[ ] \in [-1, 1] : 0 \leq 1 - (\outer[ ] + \inner[ ])} \outer[ ]^2 + \inner[ ] + 1$.
The Stackelberg equilibrium of this game is given by $\outer[ ]^* = \nicefrac{1}{2}, \inner[ ]^* = \nicefrac{1}{2}$.

If both players employ no-regret algorithms that generate strategies $\{\outer[][\iter], \inner[][\iter] \}_{\iter \in \N_+}$,
then at time $\numiters \in \N_+$, there exists $\varepsilon > 0$, s.t.
\begin{align*}\left\{
\begin{array}{c}
    \frac{1}{\numiters}\sum_{\iter = 1}^\numiters \left[{\outer[ ][\iter]}^2 + \inner[ ][\iter] + 1 \right]- \frac{1}{\numiters} \min_{\outer[ ] \in [-1, 1]} \sum_{\iter =1}^\numiters \left[\outer[ ]^2 + \inner[ ][\iter] + 1 \right] \leq \varepsilon \\
    \frac{1}{\numiters} \max_{\inner[ ] \in [-1, 1]} \sum_{\iter = 1}^\numiters \left[{\outer[ ][\iter]}^2 + \inner[ ] + 1 \right] -  \frac{1}{\numiters}\sum_{\iter = 1}^\numiters \left[{\outer[ ][\iter]}^2 + \inner[ ][\iter] + 1 \right] \leq \varepsilon  
\end{array}\right.
\end{align*}

\noindent
Simplifying yields:
\begin{align*}
\left\{
\begin{array}{c}
    \frac{1}{\numiters}\sum_{\iter = 1}^\numiters {\outer[ ][\iter]}^2 - \min_{\outer[ ] \in [-1, 1]} \outer[ ]^2 \leq \varepsilon   \\
    \max_{\inner[ ] \in [-1, 1]} \inner[ ] -  \frac{1}{\numiters}\sum_{\iter = 1}^\numiters  \inner[ ][\iter] \leq \varepsilon  
\end{array}\right.
=\left\{
\begin{array}{c}
    \frac{1}{\numiters}\sum_{\iter = 1}^\numiters {\outer[ ][\iter]}^2 \leq \varepsilon \\
    1 - \varepsilon \leq \frac{1}{\numiters}\sum_{\iter = 1}^\numiters  \inner[ ][\iter]
\end{array}\right.
\end{align*}

\noindent
In other words, the average iterates converge to $\outer[ ] = 0$, $\inner[ ] = 1$, which is not the Stackelberg equilibrium of this game.
\end{example}


If the inner player minimizes their vanilla regret without regard to the game's constraints, then their strategies are not guaranteed to be feasible, and thus cannot converge to a Stackelberg equilibrium.
To remedy this infeasibility,
we introduce a new type of regret we call \mydef{Lagrangian regret}, and show that assuming access to a solution oracle for the optimal KKT multipliers of the game's constraints, if both players minimize their Lagrangian regret, then no-regret learning dynamics converge to a Stackelberg equilibrium.

Let $\lang[\outer](\inner, \langmult) = \obj(\outer, \inner) + \sum_{\numconstr = 1}^\numconstrs \langmult[\numconstr] \constr[\numconstr](\outer, \inner)$ denote the Lagrangian associated with the outer player's value function, or equivalently, the inner player's maximization problem, given the outer player's strategy $\outer \in \outerset$.
Using this notation, we can re-express the Stackelberg game as
$\min_{\outer \in \outerset} \max_{\inner \in \innerset : \constr(\outer, \inner) \geq \zeros} \obj(\outer, \inner) = 
\min_{\outer \in \outerset} \max_{\inner \in \innerset } \min_{\langmult \geq \zeros} \\ \lang[\outer]( \inner, \langmult)$.
If the optimal KKT multipliers $\langmult^* \in \R^\numconstrs$, which are guaranteed to exist 
by Slater's condition \cite{slater1959convex}, were known, then one could plug them back into the Lagrangian to obtain a convex-concave saddle point problem given by $\min_{\outer \in \outerset} \max_{\inner \in \innerset } \lang[\outer]( \inner, \langmult^*)$.
Note that a saddle point of this problem is guaranteed to exist by the minimax theorem \cite{neumann1928theorie}, since $\lang[\outer]( \inner, \langmult^*)$ is convex in $\outer$ and concave in $\inner$.
%
The next lemma states that the Stackelberg equilibria of a min-max Stackelberg game correspond to the saddle points of $\lang[\outer](\inner, \langmult^*)$.

\begin{lemma}
\label{thm:stackelberg-equiv}
Any Stackelberg equilibrium $(\outer^* \inner^*) \in \outerset \times \innerset$ of any min-max Stackelberg game 
$(\outerset, \innerset, \obj, \constr)$ corresponds to a saddle point of $\lang[\outer](\inner, \langmult^*)$, where $\langmult^* \in \argmin_{\langmult \geq 0} \min_{\outer \in \outerset} \max_{\inner \in \innerset} \lang[\outer](\inner, \langmult)$.

\end{lemma}

This lemma tells us that the function $\lang[\outer]( \inner, \langmult^*)$ 
represents a new loss function that enforces the game's constraints.
Based on this observation, we assume access to a Lagrangian solution oracle that provides us with $\langmult^* \in \argmin_{\langmult \geq 0} \min_{\outer \in \outerset} \max_{\inner \in \innerset} \lang[\outer](\inner, \langmult^*)$.

Next, we define a new type of regret which we call \mydef{Lagrangian regret}.
Given a sequence of strategies $\left\{\outer[][\iter], \inner[][\iter]\right\}$ played by the outer and inner players in an online min-max Stackelberg game $\left\{ \left( \outerset, \innerset, \obj[\iter], \constr[][\iter] \right) \right\}$, we define their Lagrangian regret, respectively, as $\langregret[\outerset][\numiters] \left( \left\{ \outer[][\iter] \right\}, \outer \right) = \frac{1}{\numiters}\sum_{\iter = 1}^\numiters \lang[{\outer[ ][\iter]}][\iter](\inner[][\iter], \langmult^*) - \frac{1}{\numiters} \sum_{\iter =1}^\numiters  \lang[\outer][\iter] (\inner[][\iter],\langmult^*)$ and $\langregret[\innerset][\numiters] \left( \left\{ \inner[][\iter] \right\}, \inner \right) = \frac{1}{\numiters} \sum_{\iter =1}^\numiters \lang[{\outer[][\iter]}][\iter](\inner, \langmult^*) -  \frac{1}{\numiters}\sum_{\iter = 1}^\numiters \lang[{\outer[][\iter]}][\iter](\inner[][\iter], \langmult^*)$. 
We further define $\langregret[\outerset][\numiters] \left( \left\{ \outer[][\iter] \right\}\right)$ and $\langregret[\innerset][\numiters] \left( \left\{ \inner[][\iter] \right\}\right)$ as expected.

The \mydef{saddle point residual} of a point $(\outer^*, \inner^*) \in \outerset \times \innerset$ w.r.t.{} a convex-concave function $h: \outerset \times \innerset \to \R$ is given by $\max_{\inner \in \innerset} h(\outer^*, \inner) - \min_{\outer \in \outerset} h(\outer, \inner^*)$.
When the saddle point residual of $(\outer, \inner)$ w.r.t. $\lang[\outer](\inner, \langmult^*)$ is 0, 
the saddle point is a $(0, 0)$-Stackelberg equilibrium.

The main theorem of this section now follows: if both players play so as to minimize their Lagrangian regret, then their average strategies converge to a Stackelberg equilibrium. 
The bound is given in terms of the saddle point residual of the iterates generated.

\begin{theorem}
\label{thm:lang-regret-bound}
Consider a repeated min-max Stackelberg game $(\outerset, \innerset, \obj, \constr)$, and suppose the outer and the players generate sequences of strategies $\{(\outer[][\iter], \inner[][\iter])\}$ using a no-Lagrangian-regret algorithm.
If after $\numiters$ iterations, the Lagrangian regret of both players is bounded by $\varepsilon$, i.e., 
$\langregret[\outerset][\numiters] \left( \left\{ \outer[][\iter] \right\} \right) \le \varepsilon$ and
$\langregret[\innerset][\numiters] \left( \left\{ \inner[][\iter] \right\} \right) \le \epsilon$,
then the following convergence bound holds on the saddle point residual of $(\avgouter[][\numiters], \avginner[][\numiters])$ w.r.t.\ the Lagrangian:
    $0 \leq \max_{\inner \in \innerset}  \lang[{\avgouter[][\numiters]}](\inner, \langmult^*) - \min_{\outer \in \outerset}  \lang[\outer] (\avginner[][\numiters],\langmult^*) \leq 2\varepsilon$.
\end{theorem}

Having provided convergence to Stackelberg equilibrium of general no-regret learning dynamics in repeated min-max Stackelberg games, we now proceed to investigate the convergence and robustness properties of a specific example of a no-regret learning dynamic, namely online mirror descent (OMD).

\section{Online Mirror Descent}
\label{sec:omd}

In this section, we apply the results we have derived for general no-regret learning dynamics to Online Mirror Descent (OMD) specifically \cite{nemirovskij1983problem, shalev2011online}.
We then study the robustness properties of OMD in min-max Stackelberg games.

\subsection{Convergence Analysis}

When the outer player is an OMD learner minimizing its asymmetric regret and the inner player best responds, we obtain the max-oracle mirror descent (MD) algorithm (\Cref{alg:momd}), a special case of which was first proposed by \citeauthor{jin2020local} \cite{jin2020local} for min-max games (with independent strategy sets) under the name of max-oracle GD.
\citeauthor{goktas2021minmax} \cite{goktas2021minmax} extended their algorithm from min-max games (with independent strategy sets) to min-max Stackelberg games and proved its convergence in best iterates. 
Max-oracle MD (\Cref{alg:momd}) is a further generalization of both algorithms.

\begin{algorithm}[htbp]
\caption{Max-Oracle Mirror Descent (MD)}
\label{alg:momd}
\textbf{Inputs:} $\outerset, \innerset, \obj, \constr, \learnrate, \outeriters, \outer^{(0)}, \regul$ \qquad \qquad
\textbf{Output:} $\outer^{*}, \inner^{*}$
\begin{algorithmic}[1]
\For{$\outeriter = 1, \hdots, \outeriters$}
    \State Find $\inner^*(\outer[][\iter -1]) \in \br[\innerset](\outer[][\iter -1])$ 
    \State Set $\inner^{(\outeriter-1)} = \inner^*(\outer[][\iter -1])$ 
    \State Set $\langmult^{(\outeriter-1)} = \langmult^*(\outer^{(\outeriter-1)}, \inner^{(\outeriter-1)})$
    \State {\scriptsize Set $\outer[][\iter] = \argmin_{\outer \in \outerset}   \left< \grad[\outer] \lang[\outer^{(\iter-1)}]\left( \inner^{(\outeriter-1)}, \langmult^{(\outeriter-1)}\right) , \outer \right> + \frac{1}{2\learnrate[\iter]} \bregman[\regul](\outer || \outer^{(\iter-1)})$}
\EndFor
\State Set $\avgouter[][\numiters] = \frac{1}{\numiters} \sum_{\iter = 1}^\numiters \outer[][\iter]$
\State Set $\inner^*(\avgouter[][\numiters]) \in \br[\innerset](\avgouter[][\numiters])$
\State \Return $(\avgouter[][\numiters], \inner^*(\avgouter[][\numiters]))$
\end{algorithmic}
\end{algorithm}


The following corollary of \Cref{thm:pes-regret-bound}, which concerns convergence of the more general max-oracle MD algorithm in average iterates, complements \citeauthor{goktas2021minmax}'s result on the convergence of max-oracle GD (\Cref{alg:mogd}, \Cref{sec-app:algos}) in best iterates:
if the outer player employs a strategy that achieves $\varepsilon$-asymmetric regret, then the max-oracle MD algorithm is guaranteed to converge to the outer player's $(\varepsilon, 0)$-Stackelberg equilibrium strategy in average iterates after $O(\nicefrac{1}{\varepsilon^2})$ iterations, assuming the inner player best responds.

We note that 
since $\val[\outerset]$ is convex, by \Cref{thm:convex-value-func}, $\val[\outerset]$ is subdifferentiable.
Moreover, for all $\widehat{\outer} \in \outerset$, $\widehat{\inner} \in \br[\innerset](\widehat{\outer})$, $\grad[\outer] \obj(\widehat{\outer}, \widehat{\inner}) + \sum_{\numconstr = 1}^\numconstrs \langmult[\numconstr]^* \constr[\numconstr](\widehat{\outer}, \widehat{\inner})$ is an arbitrary subgradient of the value function at $\widehat{\outer}$ by \citeauthor{goktas2021minmax}'s  subdifferential envelope theorem \cite{goktas2021minmax}.
We add that similar to \citeauthor{goktas2021minmax}, we assume that the optimal KKT multipliers $\langmult^*(\outer^{(\outeriter)}, \widehat{\inner}(\outer^{(\outeriter)}))$ associated with a solution $\widehat{\inner}(\outer^{(\outeriter)}))$ can be computed in constant time.

\begin{corollary}
\label{corr:max-oracle-gradient-descent}
Let $c = \max_{\outer \in \outerset} \left\| \outer \right\|$ and let $\lipschitz[\obj] =  \max_{(\widehat{\outer}, \widehat{\inner}) \in \outerset \times \innerset} \\ \left\| \grad[\outer] \obj (\widehat{\outer}, \widehat{\inner}) \right\|$.
If \Cref{alg:momd} is run on a repeated min-max Stackelberg game $(\outerset, \innerset, \obj, \constr)$, with $\learnrate[\iter] = \frac{c}{\lipschitz[\obj] \sqrt{2T}}$, for all iteration $\iter \in \iters$ and any $\outer[][0] \in \outerset$, then $(\avgouter[][\numiters], \inner^*(\avgouter[][\numiters]))$ is a $(\nicefrac{c \lipschitz[\obj] \sqrt{2}}{\sqrt{\numiters}}, 0)$-Stackelberg equilibrium.
Furthermore, for any $\varepsilon \in (0,1)$, there exists $N(\varepsilon) \in O(\nicefrac{1}{\varepsilon^{2}})$ s.t.{} for all $\numiters \geq N(\varepsilon)$, there exists an iteration $\numiters^{*} \leq \outeriters$ s.t.\ $(\avgouter[][\numiters], \inner^*(\avgouter[][\numiters]))$ is an $(\varepsilon, 0)$-Stackelberg equilibrium.
\end{corollary}

Note that we can relax \Cref{thm:pes-regret-bound} to instead work with an approximate best response of the inner player, i.e., given the strategy of the outer player $\widehat{\outer}$, instead of playing an exact best-response, the inner player could compute a $\widehat{\inner}$ s.t.\ $\obj(\widehat{\outer}, \widehat{\inner}) \geq \max_{\inner \in \innerset : \constr(\widehat{\outer}, \inner) \geq \zeros } \obj(\widehat{\outer}) - \varepsilon$.
Moreover, the inner player could run gradient (or mirror) ascent on $\obj(\widehat{\outer}, \inner)$ to find $\widehat{\inner}$, instead of assuming a best-response oracle in \Cref{alg:momd}.
We can combine the fact that gradient ascent on Lipschitz smooth functions converges in $O(\nicefrac{1}{\varepsilon})$ iterations \cite{nemirovskij1983problem} with our novel convergence rate for max-oracle MD to conclude that the average iterates computed by nested GDA \cite{goktas2021minmax} 
converge to an $(\varepsilon, \varepsilon)$-Stackelberg equilibrium in $O(\nicefrac{1}{\varepsilon^{3}})$ iterations.
If additionally, $\obj$ is strongly convex in $\inner$, then the iteration complexity can be reduced to $O(\nicefrac{1}{\varepsilon^{2}}\log(\nicefrac{1}{\varepsilon}))$. 

Similarly, we can also consider the {symmetric} case, in which both the outer and inner players minimize their Lagrangian regrets, as OMD learners with access to a Lagrangian solution oracle that returns $\langmult^* \in \argmin_{\langmult \geq 0} \min_{\outer \in \outerset} \max_{\inner \in \innerset} \lang[\outer](\inner, \langmult)$.
In this case, we obtain the \mydef{Lagrangian mirror descent ascent (LMDA)} algorithm (Algorithm~\ref{alg:lmda}).
The following corollary of \Cref{thm:lang-regret-bound} states that LMDA
converges in average iterates to an $\varepsilon$-Stackelberg equilibrium in $O(\nicefrac{1}{\varepsilon^{2}})$ iterations.

\begin{algorithm}[htbp]
\caption{Lagrangian Mirror Descent Ascent (LMDA)}
\label{alg:lmda}
\textbf{Inputs:} $\langmult^*, \outerset, \innerset, \obj, \constr,  \learnrate[][\outer], \learnrate[][\inner], \numiters,  \outer^{(0)},  \inner^{(0)}, \regul$ \qquad
\textbf{Output:} $\outer^{*}, \inner^{*}$
\begin{algorithmic}[1]
\For{$\iter = 1, \hdots, \numiters -1$}    
    
    
    \State {\scriptsize Set $\outer[][\iter] = \argmin_{\outer \in \outerset}   \left< \grad[\outer] \lang[\outer^{(\iter-1)}]\left( \inner^{(\outeriter-1)}, \langmult^*\right) , \outer \right> + \frac{1}{2\learnrate[\iter]} \bregman[\regul](\outer || \outer^{(\iter)})$}

    
    \State {\scriptsize Set $\inner[][\iter] = \argmax_{\inner \in \innerset}   \left< \grad[\inner] \lang[\outer^{(\iter-1)}]\left( \inner^{(\iter-1)}, \langmult^*\right) , \inner \right> - \frac{1}{2\learnrate[\iter]} \bregman[\regul](\inner || \inner^{(\iter-1)})$}
    
    
\EndFor
\State \Return $\{(\outer[][\iter], \inner[][\iter])\}_{\iter= 1}^\numiters$
\end{algorithmic}
\end{algorithm}

\begin{corollary}
\label{cor:simu-omd}
Let $b = \max_{\outer \in \outerset} \left\| \outer \right\|$, $c = \max_{\inner \in \innerset} \left\| \inner \right\|$, and $\lipschitz[\lang] = \max_{(\widehat{\outer}, \widehat{\inner}) \in \outerset \times \innerset} \left\| \grad[\outer] \lang[{\widehat{\outer}}](\widehat{\inner}, \langmult^*) \right\|$.
 If \Cref{alg:lmda} is run on a repeated min-max Stackelberg game $(\outerset, \innerset, \obj, \constr)$, with $\learnrate[\iter][\outer] = \frac{b }{\lipschitz[\lang] \sqrt{2T}}$ and $\learnrate[\iter][\inner] = \frac{c }{\lipschitz[\lang] \sqrt{2T}}$, for all iterations $\iter \in \iters$ and any $\outer[][0] \in \outerset$, 
 then the following convergence bound holds on the saddle point residual of $(\avgouter[][\numiters], \avginner[][\numiters])$ w.r.t.\ the Lagrangian:
    $0 \leq \max_{\inner \in \innerset}  \lang[{\avgouter[][\numiters]}](\inner, \langmult^*) - \min_{\outer \in \outerset}  \lang[\outer] (\avginner[][\numiters],\langmult^*) \leq \frac{ 2\sqrt{2} \lipschitz[\lang]  }{\sqrt{\numiters}} \max\left\{ b, c\right\}$.
\end{corollary}

We remark that in certain rare cases the Lagrangian can become degenerate in $\inner$, in that the $\inner$ terms in the Lagrangian might cancel out when $\langmult^*$ is plugged back into Lagrangian, leading LMDA to not update the $\inner$ variables, as demonstrated by the following example:

\begin{example}
Consider the following min-max Stackelberg game:
$\min_{\outer[ ] \in [-1, 1]} \max_{\inner[ ] \in [-1, 1] : 0 \leq 1 - (\outer[ ] + \inner[ ])} \outer[ ]^2 + \inner[ ] + 1 $.
When we plug the optimal KKT multiplier $\langmult[ ]^* = 1$ into the Lagrangian associated with the outer player's value function, we obtain $\lang[{\outer[ ]}]( \inner[ ], \langmult[ ]) = \outer[ ]^2 + \inner[ ] +  1 - (\outer[ ] + \inner[ ]) = \outer[ ]^2 - \outer[ ] +  1$, with
$\frac{\partial \lang}{\partial \outer[ ]} = 2x - 1$ and $\frac{\partial \lang}{\partial \inner[ ]} = 0$.
It follows that the $\outer$ iterate converges to $\nicefrac{1}{2}$, but the $\inner$ iterate will never be updated, and hence unless $\inner$ is initialized at its Stackelberg equilibrium value, LMDA will not converge to a Stackelberg equilibrium.
\end{example}

In general, this degeneracy issue occurs when $\forall \outer \in \outerset, \grad[\inner] \obj(\outer, \inner) = - \sum_{\numconstr = 1}^\numconstrs \langmult[\numconstr]^* \grad[\inner] \constr[\numconstr](\outer, \inner)$.
We can sidestep the issue by restricting our attention to min-max Stackelberg games with convex-\emph{strictly}-concave objective functions, which is \emph{sufficient} to ensure that the Lagrangian is not degenerate in $\inner$ \cite{boyd2004convex}.
However, we observe in our experiments 
that even for convex-non-strictly-concave min-max Stackelberg games, LMDA, specifically with regularizer $\regul(\outer) = \left\| \outer\right\|_2^2$ (i.e., LGDA; \Cref{alg:lgda}, \Cref{sec-app:algos}), converges to Stackelberg equilibrium.

\subsection{Robustness Analysis}
\label{sec:robustness}

Our analysis thus far of min-max Stackelberg games has assumed the same game is played repeatedly.
In this section, we expand our consideration to 
online min-max Stackelberg games more generally, allowing the objective function to change from one time step to the next, as in the OCO framework.
Providing dynamics that are robust to ongoing game changes is crucial, as the real world is rarely static.


Online games bring with them a host of interesting issues.
Notably, even though the environment might change from one time step to the next, the game still exhibits a Stackelberg equilibrium during each stage of the game.
However, one cannot reasonably expect the players to play an equilibrium during each stage, since even in a repeated game setting, known game dynamics require multiple iterations before players can reach an approximate equilibrium.
Players cannot immediately best respond, but they can behave like boundedly rational agents who take a step in the direction of their optimal strategy during each iteration.
In general online games, equilibria also become dynamic objects, which can never be reached unless the game stops changing.


Corollaries~\ref{corr:max-oracle-gradient-descent} and ~\ref{cor:simu-omd} tell us that OMD dynamics are effective equilibrium-finding strategies in repeated min-max Stackelberg games.
However, they do not provide any intuition about the robustness of OMD dynamics to perturbations in the game.
In this section, we ask whether OMD dynamics can track Stackelberg equilibria when the game changes.
%
Ultimately, our theoretical results only concern online min-max games (with independent strategy sets), for which Nash, not Stackelberg, equilibrium is the relevant solution concept.
Nonetheless, we provide experimental evidence that suggests that the results we prove may also apply more broadly to online min-max Stackelberg games (with dependent strategy sets).
We note that our our robustness analysis focuses on projected OGD dynamics, a special case of OMD dynamics, for ease of analysis.

We first consider the asymmetric setting, in which the outer player is a no-regret learner and the inner player best-responds.
In this setting, we show that when the outer player plays according to projected OGD dynamics in an arbitrary online min-max game, the outer player's strategies closely track their Nash equilibrium strategies.
The following result states that regardless of the initial strategy of the outer player, projected OGD dynamics are always within a $\nicefrac{2d}{\delta}$ radius of the outer player's Nash equilibrium strategy.

\begin{theorem}
\label{thm:robustness_gd}
Consider an online min-max game $\left\{(\outerset, \innerset, \obj[\iter]) \right\}_{\iter = 1}^\numiters$.
Suppose that, for all $\iter \in \iters$, $\obj[\iter]$ is $\mu$-strongly convex in $\outer$ and strictly concave in $\inner$, and $ \obj[\iter]$ is $\lipschitz[{\grad\obj}]$-Lipschitz smooth.
Suppose the outer player generates a sequence of actions $\{\outer[][\iter]\}_{\iter =1}^\numiters$ by using projected OGD on the loss functions $\{ \val[][\iter]\}_{\iter = 1}^\numiters$ with learning rate $\learnrate[ ] \leq \frac{2}{\mu + \lipschitz[{\grad\obj}]}$, and further suppose the inner player generates a sequence of best-responses $\{\inner[][\iter]\}_{\iter =1}^\numiters$ to each iterate of the outer player.
For all $\iter \in \iters$, let ${\outer[][\iter]}^* \in \argmin_{\outer \in \outerset} \val[][\iter](\outer) $, $\Delta^{(\iter)} = \left\|{\outer[][\iter +1]}^* -{\outer[][\iter]}^* \right\|$, and $\delta = \frac{2 \learnrate[ ] \mu  \lipschitz[{\grad\obj}] }{\lipschitz[{\grad\obj}] + \mu}$.
We then have:
$\left\|{\outer[][\numiters]}^* - \outer[][\numiters]\right\|    \leq (1 - \delta)^{\nicefrac{\numiters}{2}} \left\|{\outer[][0]}^* - \outer[][0]\right\| + \sum_{\iter = 1}^\numiters \left( 1 - \delta \right)^{\frac{\numiters - \iter}{2}} \Delta^{(\iter)}$.
%
If additionally, for all $\iter \in \iters$, $\Delta^{(\iter)} \leq d$, then:
$\left\|{\outer[][\numiters]}^* - \outer[][\numiters]\right\|    \leq (1 - \delta)^{\nicefrac{\numiters}{2}} \left\|{\outer[][0]}^* - \outer[][0]\right\| + \frac{2d}{\delta}$.
\end{theorem}

We can derive a similar robustness result in the symmetric setting, where the outer and inner players are both projected OGD learners.
The following result states that regardless of the initial strategies of the two players, projected OGD dynamics follow the Nash equilibrium of the game, always staying within a $\nicefrac{4d}{\delta}$ radius.

\begin{theorem}
\label{thm:robustness_lgda}
Consider an online min-max game $
\left\{(\outerset, \innerset, \obj[\iter]) \right\}_{\iter = 1}^\numiters$.
Suppose that, for all $\iter \in \iters$, $\obj[\iter]$ is $\mu_\outer$-strongly convex in $\outer$ and $\mu_\inner$-strongly concave in $\inner$, and $\obj[\iter]$ is $\lipschitz[{ \grad \obj}]$-Lipschitz smooth.
Let $\{(\outer[][\iter], \inner[][\iter])\}_{\iter =1}^\numiters$ be the strategies played by the outer and inner players, assuming that the outer player uses a projected OGD algorithm on the losses $\{ \obj[\iter](\cdot, \inner[][\iter])\}_{\iter =1}^\numiters$ with $\learnrate[\outer] = \frac{2}{\mu_\outer + \lipschitz[{\grad \obj}]}$ and the inner player uses a projected OGD algorithm on the losses $\{ - \obj[\iter](\outer[][\iter], \cdot)\}_{\iter =1}^\numiters$ with $\learnrate[\inner] = \frac{2}{\mu_\inner + \lipschitz[{\grad \obj}]}$.
For all $\iter \in \iters$, let ${\outer[][\iter]}^* \in \argmin_{\outer \in \outerset} \obj[\iter](\outer, \inner[][\iter]) $, ${\inner[][\iter]}^* \in \argmin_{\inner \in \innerset} \obj[\iter](\outer[][\iter], \inner)$, $\Delta^{(\iter)}_{\outer} = \left\|{\outer[][\iter +1]}^* -{\outer[][\iter]}^* \right\|$, $\Delta^{(\iter)}_{\inner} = \left\|{\inner[][\iter +1]}^* -{\inner[][\iter]}^* \right\|$, $\delta_\outer = \frac{2 \learnrate[ ] \mu_\outer \lipschitz[{\grad\obj}] }{\lipschitz[{\grad[\outer] \obj}] + \mu_\outer}$, and $\delta_\inner = \frac{2 \learnrate[ ] \mu_\inner \lipschitz[{\grad\obj}] }{\lipschitz[{\grad\obj}] + \mu_\inner}$.
We then have:
%
$\left\|{\outer[][\numiters]}^* - \outer[][\numiters]\right\| + \left\|{\inner[][\numiters]}^* - \inner[][\numiters]\right\| 
\leq (1 - \delta_\outer)^{\nicefrac{\numiters}{2}} \left\|{\outer[][0]}^* - \outer[][0]\right\| + (1 - \delta_\inner)^{\nicefrac{\numiters}{2}} \left\|{\inner[][0]}^* - \inner[][0]\right\| 
+ \sum_{\iter = 1}^\numiters \left( 1 - \delta_\outer \right)^{\frac{\numiters - \iter}{2}} \Delta_\outer^{(\iter)} + \sum_{\iter = 1}^\numiters \left( 1 - \delta_\inner \right)^{\frac{\numiters - \iter}{2}} \Delta_\inner^{(\iter)}$.
%
If additionally, for all $\iter \in \iters$, $\Delta_\outer^{(\iter)} \leq d$ and $\Delta_\inner^{(\iter)} \leq d$, and $\delta = \min\{\delta_\inner, \delta_\outer\}$, then:
%
$\left\|{\outer[][\numiters]}^* - \outer[][\numiters]\right\|  + \left\|{\inner[][\numiters]}^* - \inner[][\numiters]\right\|
\leq 2(1 - \delta)^{\nicefrac{\numiters}{2}} \\
\left(  \left\|{\outer[][0]}^* - \outer[][0]\right\| + \left\|{\inner[][0]}^* - \inner[][0]\right\| \right) + \frac{4d}{\delta}$.
\end{theorem}

The proofs of the above theorems are relegated to \Cref{sec_app:proofs}.
These theorems establish the robustness of projected OGD dynamics for min-max games in both the asymmetric and symmetric settings by showing that the dynamics closely track the Nash equilibria in a large class of min-max games (with independent strategy sets). These results also suggest that general OMD dynamics, e.g., OMD with entropy as a regularizer, are robust to perturbation.
As we are not able to extend these theoretical robustness guarantees to min-max Stackelberg games (with dependent strategy sets), we instead ran a series of experiments with online Fisher markets, which are canonical examples of min-max Stackelberg games \cite{goktas2021minmax}, to investigate the empirical robustness guarantees of projected OGD dynamics for this class of min-max Stackelberg games.

\section{Online Fisher Markets}
\label{sec:experiments}

The Fisher market model, attributed to Irving Fisher \cite{brainard2000compute}, has received a great deal of attention in the literature, especially by computer scientists, as it has proven useful in the design of electronic marketplaces.
We now study OMD dynamics in online Fisher markets, which are instances of min-max Stackelberg games \cite{goktas2021minmax}.

A \mydef{Fisher market} consists of $\numbuyers$ buyers and $\numgoods$ divisible goods \cite{brainard2000compute}.
Each buyer $\buyer \in \buyers$ has a budget $\budget[\buyer] \in \mathbb{R}_{+}$ and a utility function $\util[\buyer]: \mathbb{R}_{+}^{\numgoods} \to \mathbb{R}$.
Each good $\good \in \goods$ has supply $\supply[\good] \in \R_+$.
A Fisher market is thus given by a tuple $(\numbuyers, \numgoods, \util, \budget, \supply)$, where $\util = \left\{\util[1], \hdots, \util[\numbuyers] \right\}$ is a set of utility functions, one per buyer; $\budget \in \R_{+}^{\numbuyers}$ is a vector of buyer budgets; and $\supply \in \R^\numgoods_+$ is a vector of good supplies.
We abbreviate as $(\util, \budget, \supply)$ when $\numbuyers$ and $\numgoods$ are clear from context.
An \mydef{online Fisher market} is a sequence of Fisher markets $\left\{\left( \util^{(\iter)}, \budget^{(\iter)}, \supply^{(\iter)} \right)\right\}_{\iter = 1}^{\numiters}$.

An \mydef{allocation} $\allocation = \left(\allocation[1], \hdots, \allocation[\numbuyers] \right)^T \in \R_+^{\numbuyers \times \numgoods}$ is an assignment of goods to buyers, represented as a matrix s.t.\ $\allocation[\buyer][\good] \ge 0$ denotes the amount of good $\good \in \goods$ allocated to buyer $\buyer \in \buyers$.
Goods are assigned \mydef{prices} $\price = \left(\price[1], \hdots, \price[\numgoods] \right)^T \in \mathbb{R}_+^{\numgoods}$.
A tuple $(\price^*, \allocation^*)$ is said to be a \mydef{competitive 
equilibrium (CE)} of Fisher market $(\util, \budget, \supply)$ if 1.~buyers are utility maximizing, constrained by their budget, i.e., $\forall \buyer \in \buyers, \allocation[\buyer]^* \in \argmax_{\allocation[ ] : \allocation[ ] \cdot \price^* \leq \budget[\buyer]} \util[\buyer](\allocation[ ])$;
and 2.~the market clears, i.e., $\forall \good \in \goods,  \price[\good]^* > 0 \Rightarrow \sum_{\buyer \in \buyers} \allocation[\buyer][\good]^* = \supply[\good]$  and $\price[\good]^* = 0 \Rightarrow\sum_{\buyer \in \buyers} \allocation[\buyer][\good]^* \leq \supply[\good]$.

\citeauthor{goktas2021minmax} \cite{goktas2021minmax} observe that any CE $(\price^*, \allocation^*)$ of a Fisher market $(\util, \budget)$ corresponds to a Stackelberg equilibrium of the following min-max Stackelberg game:%
\footnote{The first term in this program is slightly different than the first term in the program presented by \citeauthor{goktas2021minmax} \cite{goktas2021minmax}, since supply is assumed to be 1 their work.}
\begin{align}
    \min_{\price \in \R_+^\numgoods} \max_{\allocation \in \R^{\numbuyers \times \numgoods}_+ :  \allocation \price \leq \budget} \sum_{\good \in \goods} \supply[\good] \price[\good] + \sum_{\buyer \in \buyers}  \budget[\buyer] \log \left(  \util[\buyer](\allocation[\buyer]) \right) \enspace .
    \label{fisher-program}
\end{align}

\noindent
Let $\lang: \R^\numgoods_+ \times \R^{\numbuyers \times \numgoods} \to \R_+$ be the Lagrangian of the outer player's value function in \Cref{fisher-program}, i.e., 
$\lang[\price](\allocation, \langmult) = \sum_{\good \in \goods} \supply[\good] \price[\good] \\ + \sum_{\buyer \in \buyers}  \budget[\buyer] \log \left(  \util[\buyer](\allocation[\buyer]) \right) + \sum_{\buyer \in \buyers} \langmult[\buyer] \left( \budget[\buyer] - \allocation[\buyer] \cdot \price \right)$. One can show the existence of a Lagrangian solution oracle for the Lagrangian of \Cref{fisher-program} such that $\langmult^* = \ones[\numgoods]$. 
We then have: 1.~ by \citeauthor{goktas2021minmax}'s envelope theorem, the subdifferential of the outer player's value function is given by $\grad[\price] \val(\price) = \supply - \sum_{\buyer \in \buyers} \allocation[\buyer]^*(\price)$, where $\allocation[\buyer]^*(\price) \in \argmax_{\allocation[ ] \in \R^\numgoods_+ \allocation[ ] \cdot \price \leq \budget[\buyer]} \util[\buyer](\allocation[ ])$, 2.~the gradient of the Lagrangian w.r.t. the prices, given the Lagrangian solution oracle, is $\grad[\price] \lang[\price](\allocation, \langmult^*) = \supply - \sum_{\buyer \in \buyers} \allocation[\buyer]$  
and
$\grad[{\allocation[\buyer]}] \lang[\price](\allocation, \langmult^*)) = \frac{\budget[\buyer]}{\util[\buyer]\left(\allocation[\buyer]\right)} \grad[{\allocation[\buyer]}] \util[\buyer]\left(\allocation[\buyer]\right) - \price$, 
where $\langmult^* = \ones[\numgoods]$ \cite{goktas2021consumer}. 


We first consider OMD dynamics for Fisher markets in the asymmetric setting, in which the outer player determines their strategy via projected OGD {first} and the inner player best-responds. 
This setup yields a dynamic version of a natural price adjustment process known as t\^atonnement \cite{walras}, this variant of which
was first studied by \citeauthor{cheung2019tracing} \cite{cheung2019tracing} (\Cref{alg:dynamic_max_oracle_gd}, \Cref{sec-app:algos}). 


We also consider OMD dynamics in the {symmetric} setting, specifically the case in which both the outer and inner players employ projected OGD {simultaneously}, which yields myopic best-response dynamics \cite{monderer1996potential} (\Cref{alg:dynamic_lgda}, \Cref{sec-app:algos}).
In words, 
at each time step, the (fictional Walrasian) auctioneer takes a gradient descent step to minimize its regret, and then all the buyers take a gradient ascent step to minimize their Lagrangian regret.
These GDA dynamics can be seen as myopic best-response dynamics for boundedly rational 
sellers and buyers.


\paragraph{Experiments}

In order to better understand the robustness properties of Algorithms~\ref{alg:dynamic_max_oracle_gd} and~\ref{alg:dynamic_lgda} in an {online} min-max Stackelberg game that is subject to perturbation across time, we ran a series of experiments with {online} Fisher Markets assuming three different classes of utility functions.%
\footnote{Our code can be found at \coderepo.}
Each utility structure endows \Cref{fisher-program} with different smoothness properties, which allows us to compare the efficiency of the algorithms under varying conditions.
Let $\valuation[\buyer] \in \R^\numgoods$ be a vector of valuation parameters that describes the utility function of buyer $\buyer \in \buyers$.
We consider the following utility function classes:
1.~linear: $\util[\buyer](\allocation[\buyer]) = \sum_{\good \in \goods} \valuation[\buyer][\good] \allocation[\buyer][\good]$; 2.~Cobb-Douglas:  $\util[\buyer](\allocation[\buyer]) = \prod_{\good \in \goods} \allocation[\buyer][\good]^{\valuation[\buyer][\good]}$; and 3.~Leontief:  $\util[\buyer](\allocation[\buyer]) = \min_{\good \in \goods} \left\{ \frac{\allocation[\buyer][\good]}{\valuation[\buyer][\good]}\right\}$.

To simulate an {online} Fisher market, we fix a range for every market parameter and draw from that range uniformly at random during each iteration.
Our goal is to understand how closely OMD dynamics track the CE of the Fisher markets as they vary with time.
We compare the iterates $\left(\price^{(\iter)}, \allocation^{(\iter)} \right)$ computed by the algorithms and the CE $\left(\price^{(\iter)^{*}}, \allocation^{(\iter)^{*}} \right)$ of the market $(\util^{(\iter)}, \budget^{(\iter)}, \supply^{(\iter)})$ at each iteration $\iter$.
The difference between these outcomes is measured as $\left\| {\price^{(\iter)^{*}} - \price^{(\iter)}} \right\|_2 + \left\| {\allocation^{(\iter)^{*}} - \allocation^{(\iter)}} \right\|_2$.

\begin{figure*}
  \begin{minipage}[c]{0.625\textwidth}
    \includegraphics[width=\textwidth]{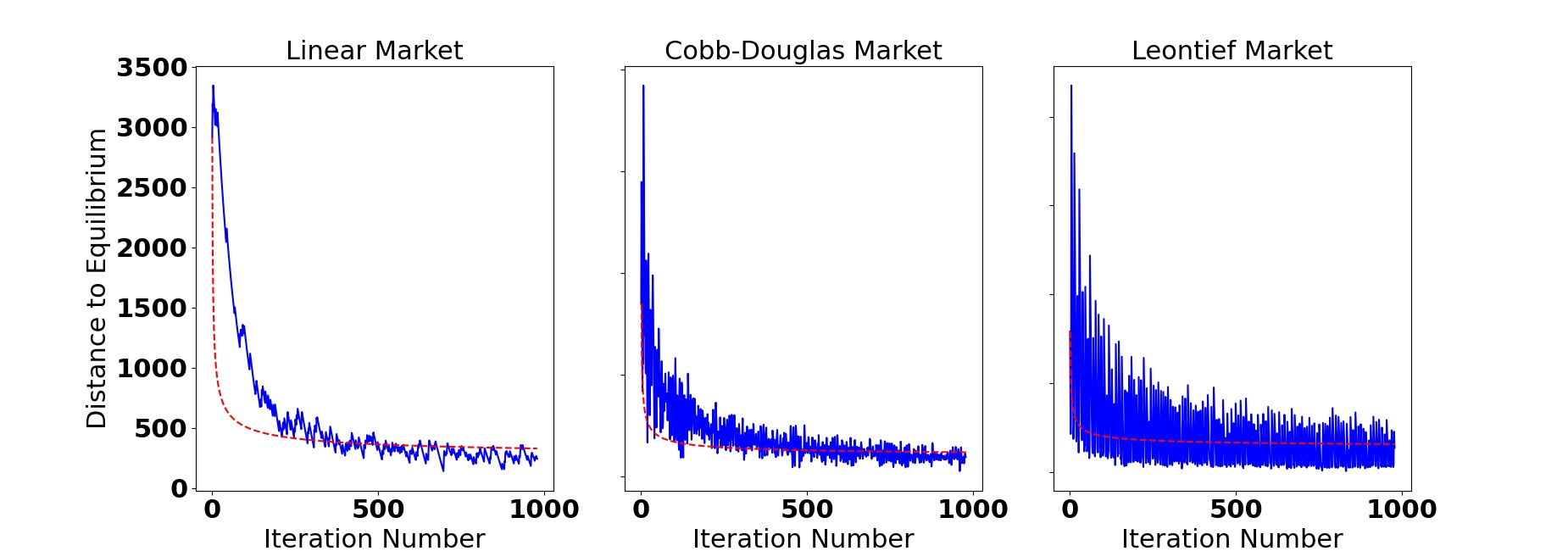}
  \end{minipage}\hfill
  \begin{minipage}[c]{0.33\textwidth}
    \caption{In {\color{blue} blue}, we depict a trajectory of distances between computed allocation-price pairs and equilibrium allocation-price pairs, when \Cref{alg:dynamic_max_oracle_gd} is run on randomly initialized online linear, Cobb-Douglas, and Leontief Fisher markets. In {\color{red} red}, we plot an arbitrary $O(\nicefrac{1}{\sqrt{T}})$ function.}
    \label{fig:exp_results_gd}
  \end{minipage}
  \begin{minipage}[c]{0.625\textwidth}
    \includegraphics[width=\textwidth]{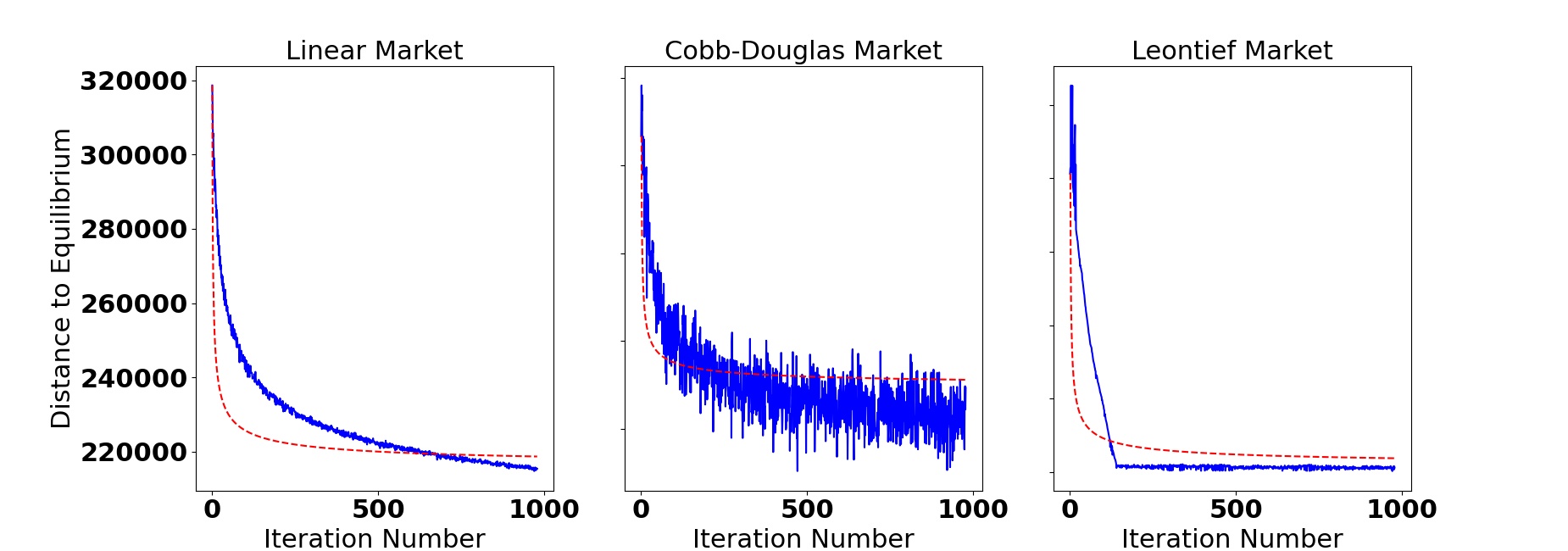}
  \end{minipage}\hfill
  \begin{minipage}[c]{0.33\textwidth}
    \caption{In {\color{blue} blue}, we depict a trajectory of distances between computed allocation-price pairs and equilibrium allocation-price pairs, when \Cref{alg:dynamic_lgda} is run on randomly initialized online linear, Cobb-Douglas, and Leontief Fisher markets. In {\color{red} red}, we plot an arbitrary $O(\nicefrac{1}{\sqrt{T}})$ function.}
    \label{fig:exp_results_lgda}
  \end{minipage}
\end{figure*}

In our experiments, we ran Algorithms~\ref{alg:dynamic_max_oracle_gd} and~\ref{alg:dynamic_lgda} on 100 randomly initialized {online} Fisher markets.
We depict the distance to the CE at each iteration for a single experiment chosen at random in Figures~\ref{fig:exp_results_gd} and~\ref{fig:exp_results_lgda}.
In these figures, we observe that the OMD dynamics are closely tracking the CE as they vary with time.
A more detailed description of our experimental setup can be found in \Cref{sec-app:fisher}.
 
We observe from Figures~\ref{fig:exp_results_gd} and~\ref{fig:exp_results_lgda} that for both Algorithms~\ref{alg:dynamic_max_oracle_gd} and~\ref{alg:dynamic_lgda}, we obtain an empirical convergence rate relatively close to $O(\nicefrac{1}{\sqrt{T}})$ under Cobb-Douglas utilities, and a slightly slower empirical convergence rate under linear utilities.
Recall that $O(\nicefrac{1}{\sqrt{T}})$ is the convergence rate guarantee we obtained for both algorithms, assuming a fixed learning rate in a repeated Fisher market  (Corollaries~\ref{corr:max-oracle-gradient-descent} and~\ref{cor:simu-omd}).
Our theoretical results assume fixed learning rates, but since those results apply to repeated games while our experiments apply to {online} Fisher markets, we selected variable learning rates. 
After manual hyper-parameter tuning, for \Cref{alg:dynamic_max_oracle_gd}, we chose a dynamic learning rate of $\learnrate[\iter][ ] = \frac{1}{\sqrt{\iter}}$, while for \Cref{alg:dynamic_lgda}, we chose learning rates of $\learnrate[\iter][\outer] = \frac{5}{\sqrt{\iter}}$ and $\learnrate[\iter][\inner] = \frac{0.01}{\sqrt{\iter}}$, for all $\iter \in \iters$.
For these optimized learning rates, we obtain empirical convergence rates close to what the theory predicts.
 
In Fisher markets with Leontief utilities, the objective function is not differentiable.
Correspondingly, {online} Fisher markets with Leontief utilities are the hardest markets of the three for our algorithms to solve. 
Still, we only see a slightly slower than $O(\nicefrac{1}{\sqrt{T}})$ empirical convergence rate.
In these experiments, the convergence curve generated by \Cref{alg:dynamic_lgda} has a less erratic behavior than the one generated by \Cref{alg:dynamic_max_oracle_gd}.
Due to the non-differentiability of the objective function, the gradient ascent step in \Cref{alg:dynamic_lgda} for buyers with Leontief utilities is very small, 
effectively dampening any potentially erratic changes in the iterates. 
 

Our experiments suggest that OMD dynamics (Algorithms~\ref{alg:dynamic_max_oracle_gd} and \ref{alg:dynamic_lgda}) are robust enough to closely track the changing CE in {online} Fisher markets.
We note that  t\^atonnement dynamics (\Cref{alg:dynamic_max_oracle_gd}) seem to be more robust than myopic best response dynamics (\Cref{alg:dynamic_lgda}), i.e., the distance to equilibrium allocations is smaller at each iteration of t\^atonnement.
This result is not surprising, as t\^atonnement computes a utility-maximizing allocation for the buyers at each time step.
Even though Theorems~\ref{thm:robustness_gd} and~\ref{thm:robustness_lgda} only provide theoretical guarantees on the robustness of OMD dynamics in online min-max games (with independent strategy sets), it seems that similar theoretical robustness results may be attainable in online min-max Stackelberg games (with dependent strategy sets). 

\section{Conclusion}

We began this paper by considering no-regret learning dynamics in repeated min-max Stackelberg games in two settings: an asymmetric setting in which the outer player is a no-regret learner and the inner player best responds, and a {symmetric} setting in which both players are no-regret learners.
For both of these settings, we proved that no-regret learning dynamics converge to a Stackelberg equilibrium of the game.
We then specialized the no-regret algorithm employed by the players to online mirror descent (OMD), which yielded two new algorithms, max-oracle MD and nested MDA in the asymmetric setting, and a new simultaneous GDA-like algorithm \cite{nedic2009gda}, which we call Lagrangian MDA, in the symmetric setting.
As these algorithms are no-regret learning algorithms, our earlier theorems imply convergence to $\varepsilon$-Stackelberg equilibria in $O(\nicefrac{1}{\varepsilon^2})$ iterations for max-oracle MD and LMDA, and $O(\nicefrac{1}{\varepsilon^3})$ iterations for nested MDA.

Finally, as many real-world applications involve changing environments, we investigated the robustness of OMD dynamics by analyzing how closely they track Stackelberg equilibria in arbitrary online min-max Stackelberg games.
We proved that in min-max games (with independent strategy sets) OMD dynamics closely track the changing Stackelberg equilibria of a game.
As we were not able to extend these theoretical robustness guarantees to min-max Stackelberg games (with dependent strategy sets), we instead ran a series of experiments with online Fisher markets, which are canonical examples of min-max Stackelberg games.
Our experiments suggest that OMD dynamics are robust for min-max Stackelberg games so that perhaps the robustness guarantees we have provided for OMD dynamics in min-max games (with independent strategy sets) can be extended to min-max Stackelberg games (with dependent strategy sets). 

The theory developed in this paper opens the door to extending the myriad applications of Stackelberg games in AI to incorporating dependent strategy sets.
Such models promise to be more expressive, and as a result could provide decision makers with better solutions to problems in security, environmental protection, etc.

\begin{acks}
We thank several anonymous reviewers for their feedback on an earlier draft of this paper.

This work was partially supported by NSF Grant CMMI-1761546.
\end{acks}



\bibliographystyle{ACM-Reference-Format} \balance
\bibliography{references.bib}


\appendix
\clearpage
\section{Additional Related Work}\label{sec-app:related}

We provide a survey of the min-max literature as presented by \citeauthor{goktas2021minmax} in what follows.  Much progress has been made recently in solving min-max games with independent strategy sets, both in the convex-concave case and in the non-convex-concave case. 
For the former case, when $\obj$ is $\mu_\outer$-strongly-convex in $\outer$ and $\mu_\inner$-strongly-concave in $\inner$, \citeauthor{tseng1995variational} \cite{tseng1995variational}, \citeauthor{nesterov2006variational} \cite{nesterov2006variational}, and \citeauthor{gidel2020variational} \cite{gidel2020variational} proposed variational inequality methods, and \citeauthor{mokhtari2020convergence} \cite{mokhtari2020convergence}, gradient-descent-ascent (GDA)-based methods, all of which compute a solution in $\tilde{O}(\mu_\inner + \mu_\outer)$ iterations.
These upper bounds were recently complemented by the lower bound of $\tilde{\Omega}(\sqrt{\mu_\inner \mu_\outer})$, shown by \citeauthor{ibrahim2019lower}  \cite{ibrahim2019lower} and  \citeauthor{zhang2020lower}   \cite{zhang2020lower}.
Subsequently, \citeauthor{lin2020near}  \cite{lin2020near} and   \citeauthor{alkousa2020accelerated} \cite{alkousa2020accelerated} analyzed algorithms that converge in $\tilde{O}(\sqrt{\mu_\inner \mu_\outer})$ and $\tilde{O}(\min\left\{\mu_\outer \sqrt{\mu_\inner}, \mu_\inner \sqrt{\mu_\outer} \right\})$ iterations, respectively. 

For the special case where $\obj$ is $\mu_\outer$-strongly convex in $\outer$ and linear in $\inner$, \citeauthor{juditsky2011first}  \cite{juditsky2011first},  \citeauthor{hamedani2018primal}  \cite{hamedani2018primal}, and \citeauthor{zhao2019optimal}  \cite{zhao2019optimal} all present methods that converge to an $\varepsilon$-approximate solution in $O(\sqrt{\nicefrac{\mu_\outer}{\varepsilon}})$ iterations.
When the strong concavity or linearity assumptions of $\obj$ on $\inner$ are dropped, and
$\obj$ is assumed to be $\mu_\outer$-strongly-convex in $\outer$ but only concave in $\inner$, \citeauthor{thekumparampil2019efficient} \cite{thekumparampil2019efficient} provide an algorithm that converges to an $\varepsilon$-approximate solution in $\tilde{O}(\nicefrac{\mu_\outer}{\varepsilon})$ iterations, and \citeauthor{ouyang2018lower} \cite{ouyang2018lower} provide a lower bound of $\tilde{\Omega}\left(\sqrt{\nicefrac{\mu_\outer}{\varepsilon}}\right)$ iterations on this same computation.
\citeauthor{lin2020near} then went on to develop a faster algorithm, with iteration complexity of $\tilde{O}\left(\sqrt{\nicefrac{\mu_\outer}{\varepsilon}}\right)$, under the same conditions.

When $\obj$ is simply assumed to be convex-concave, \citeauthor{nemirovski2004prox} \cite{nemirovski2004prox}, \citeauthor{nesterov2007dual} \cite{nesterov2007dual}, and \citeauthor{tseng2008accelerated} \cite{tseng2008accelerated} describe algorithms that solve for an $\varepsilon$-approximate solution with $\tilde{O}\left(\varepsilon^{-1}\right)$ iteration complexity, and \citeauthor{ouyang2018lower} \cite{ouyang2018lower} prove a corresponding lower bound of $\Omega(\varepsilon^{-1})$.

When $\obj$ is assumed to be non-convex-$\mu_\inner$-strongly-concave, and the goal is to compute a first-order Nash, \citeauthor{sanjabi2018stoch} \cite{sanjabi2018stoch} provide an algorithm that converges to $\varepsilon$-an approximate solution in $O(\varepsilon^{-2})$ iterations.
\citeauthor{jin2020local} \cite{jin2020local}, \citeauthor{rafique2019nonconvex} \cite{rafique2019nonconvex}, \citeauthor{lin2020gradient} \cite{lin2020gradient}, and \citeauthor{lu2019block} \cite{lu2019block} provide algorithms that converge in $\tilde{O}\left(\mu_\inner^2 \varepsilon^{-2}\right)$ iterations, while \citeauthor{lin2020near} \cite{lin2020near} provide an even faster algorithm, with an iteration complexity of $\tilde{O}\left(\sqrt{\mu_\inner} \varepsilon^{-2}\right)$.

When $\obj$ is non-convex-non-concave and the goal to compute is an approximate first-order Nash equilibrium, \citeauthor{lu2019block} \cite{lu2019block} provide an algorithm with iteration complexity $\tilde{O}(\varepsilon^{-4})$, while \citeauthor{nouiehed2019solving} \cite{nouiehed2019solving} provide an algorithm with iteration complexity $\tilde{O}(\varepsilon^{-3.5})$. More recently, \citeauthor{ostrovskii2020efficient} \cite{ostrovskii2020efficient} and \citeauthor{lin2020near} \cite{lin2020near} proposed an algorithm with iteration complexity $\tilde{O}\left(\varepsilon^{-2.5}\right)$.

When $\obj$ is non-convex-non-concave and the desired solution concept is a ``local'' Stackelberg equilibrium, \citeauthor{jin2020local} \cite{jin2020local}, \citeauthor{rafique2019nonconvex} \cite{rafique2019nonconvex}, and \citeauthor{lin2020gradient} \cite{lin2020gradient} provide algorithms with a $\tilde{O}\left( \varepsilon^{-6} \right)$ complexity.
More recently, \citeauthor{thekumparampil2019efficient} \cite{thekumparampil2019efficient}, \citeauthor{zhao2020primal} \cite{zhao2020primal}, and \citeauthor{lin2020near} \cite{lin2020near} have proposed algorithms that converge to an $\varepsilon$-approximate solution in $\tilde{O}\left( \varepsilon^{-3}\right)$ iterations.

We summarize the literature pertaining to the convex-concave and the non-convex-concave settings in Tables 1 and 2
respectively.

\newpage

\renewcommand*\arraystretch{1.5}
\begin{table}[H]
    \centering
    \caption{Iteration complexities for min-max games with independent strategy sets in convex-concave settings. Note that these results assume that the objective function is Lipschitz-smooth.} \label{tab:fixed-convex-concave}
    \begin{tabular}{|p{0.15\textwidth}|p{0.15\textwidth}|p{0.13\textwidth}|}\hline
    Setting & Reference & Iteration Complexity \\ \hline
    \multirow{8}{*}{\small\shortstack{\small $\mu_\outer$-Strongly-Convex-\\ $\mu_\inner$-Strongly-Concave}} & \cite{tseng1995variational} & \multirow{4}{*}{$\tilde{O}\left( \mu_\outer + \mu_\inner\right)$} \\\cline{2-2}
         & \cite{nesterov2006variational}  & \\ \cline{2-2}
         & \cite{gidel2020variational}     & \\ \cline{2-2}
         & \cite{mokhtari2020convergence}  &  \\ \cline{2-3}
         & \cite{alkousa2020accelerated}   & \shortstack{$\tilde{O}(\min \left\{\mu_\outer \sqrt{\mu_\inner},\right.$ \\ $\left.\mu_\inner \sqrt{\mu_\outer} ] ) \right\}$}\\ \cline{2-3}
         & \cite{lin2020near}              & $\tilde{O}(\sqrt{\mu_\outer \mu_\inner})$ \\ \cline{2-3}
         & \cite{ibrahim2019lower} & $\tilde{\Omega}(\sqrt{\mu_\outer \mu_\inner})$\\ \cline{2-2}
         & \cite{zhang2020lower} & \\ \hline \hline
    \multirow{3}{*}{\small\shortstack{$\mu_\outer$-Strongly-Convex\\-Linear}}    & \cite{juditsky2011first} & \multirow{3}{*}{$O\left( \sqrt{\nicefrac{\mu_\outer}{\varepsilon}}\right)$} \\\cline{2-2}
    & \cite{hamedani2018primal} & \\\cline{2-2}
    & \cite{zhao2019optimal}& \\\hline \hline
    \multirow{3}{*}{\small\shortstack{$\mu_\outer$-Strongly-Convex\\-Concave}} & \cite{thekumparampil2019efficient} & $\tilde{O}\left( \nicefrac{\mu_\outer }{\sqrt{\varepsilon}} \right)$ \\ \cline{2-3}
    & \cite{lin2020near} & $\tilde{O}(\sqrt{\nicefrac{\mu_\outer}{\varepsilon}})$ \\ \cline{2-3}
    & \cite{ouyang2018lower} & $\tilde{\Omega}\left( \sqrt{\nicefrac{\mu_\outer}{\varepsilon}}\right)$ \\ \hline \hline
    \multirow{5}{*}{\small\shortstack{Convex\\-Concave}} & \cite{nemirovski2004prox} & \multirow{2}{*}{$O\left( \varepsilon^{-1}\right)$} \\ \cline{2-2}
    & \cite{nesterov2007dual} & \\ \cline{2-2}
    & \cite{tseng2008accelerated} & \\ \cline{2-3}
    & \cite{lin2020near} &  $\tilde{O}\left(\varepsilon^{-1}\right)$\\ \cline{2-3}
    & \cite{ouyang2018lower} & $\Omega(\varepsilon^{-1})$ \\ \hline 
    \end{tabular}
    \renewcommand*\arraystretch{1}
\end{table}

\begin{table}[H]
    \centering
    \caption{Iteration complexities for min-max games with independent strategy sets in non-convex-concave settings. Note that although all these results assume that the objective function is Lipschitz-smooth, some authors make additional assumptions: e.g., \cite{nouiehed2019solving} obtain their result for objective functions that satisfy the Lojasiwicz condition.}
    \label{tab:fixed-nonconvex-concave}
    \renewcommand*\arraystretch{1.5}
    \begin{tabular}{|p{0.1\textwidth}|p{0.2\textwidth}|p{0.1\textwidth}|}\hline
    Setting & Reference & Iteration Complexity\\ \hline
        \multirow{5}{*}{\tiny \makecell{Nonconvex-$\mu_\inner$-\\ Strongly-Concave,\\ First Order Nash  \\ or Local Stackelberg\\ Equilibrium}} & \cite{jin2020local} & \multirow{4}{*}{$ \tilde{O}(\mu_\inner^2 \varepsilon^{-2})$} \\
         & \cite{rafique2019nonconvex} & \\ \cline{2-2}
         & \cite{lin2020gradient}  & \\ \cline{2-2}
         & \cite{lu2019block} & \\ \cline{2-3}
         & \cite{lin2020near} & $\tilde{O}\left( \sqrt{\mu_\inner} \varepsilon^{-2} \right)$\\ \hline \hline
        \multirow{4}{*}{\tiny \makecell{Nonconvex-\\Concave,\\ First Order \\ Nash Equilibrium}} & \cite{lu2019block}  & $\tilde{O}\left(\varepsilon^{-4}\right)$ \\ \cline{2-3}
        & \cite{nouiehed2019solving} & $\tilde{O}\left( \varepsilon^{-3.5}\right)$ \\ \cline{2-3}
        & \cite{ostrovskii2020efficient} & \multirow{2}{*}{$\tilde{O}\left( \varepsilon^{-2.5}\right)$} \\ \cline{2-2}
        & \cite{lin2020near} &  \\ \hline \hline
        \multirow{6}{*}{\tiny  \makecell{Nonconvex-\\Concave,\\ Local Stackelberg\\ Equilibrium}} & \cite{jin2020local} & \multirow{3}{*}{$\tilde{O}(\varepsilon^{-6})$}\\  \cline{2-2}
        & \cite{nouiehed2019solving} & \\ \cline{2-2}
        & \cite{lin2020near} & \\ \cline{2-3}
        & \cite{thekumparampil2019efficient} & \multirow{3}{*}{$\tilde{O}(\varepsilon^{-3})$}\\ \cline{2-2}
        & \cite{zhao2020primal} & \\
        & \cite{lin2020near} & \\ \hline 
    \end{tabular}
    \renewcommand*\arraystretch{1}
\end{table}

\newpage
\section{Omitted Proofs}\label{sec_app:proofs}

\begin{proof}[Proof of \Cref{thm:pes-regret-bound}]
Since {asymmetric} regret is bounded by $\varepsilon$ after $\numiters$ iterations, it holds that: 
\begin{align}
    \max_{\outer \in \outerset} \pesregret[\outerset][\numiters](\outer) &\leq \varepsilon\\
    \frac{1}{\numiters} \sum_{\iter = 1}^\numiters \val[\outerset][\iter](\outer[][\iter]) - \min_{\outer \in \outerset} \sum_{\iter =1}^\numiters  \frac{1}{\numiters} \val[\outerset][\iter](\outer) &\leq \varepsilon
\end{align}

\noindent
Since the game is static, and it further holds that:
\begin{align}
    \frac{1}{\numiters} \sum_{\iter = 1}^\numiters \val[\outerset](\outer[][\iter]) - \min_{\outer \in \outerset} \sum_{\iter =1}^\numiters  \frac{1}{\numiters} \val[\outerset](\outer) &\leq \varepsilon\\
    \frac{1}{\numiters} \sum_{\iter = 1}^\numiters \val[\outerset](\outer[][\iter]) - \min_{\outer \in \outerset} \val[\outerset](\outer) &\leq \varepsilon
\end{align}

\noindent
Thus, by the convexity of $\val[\outerset]$ (see \Cref{thm:convex-value-func}),
$\val[\outerset] (\avgouter[][\numiters]) - \min_{\outer \in \outerset} \val[\outerset] (\outer) \leq \varepsilon$.
Now replacing $\val[\outerset]$ by its definition, and setting $\inner^*(\avgouter[][\numiters]) \in \br[\innerset](\avgouter[][\numiters])$, we obtain that $\left( \avgouter[][\numiters], \inner^*(\avgouter[][\numiters]) \right)$ is $(\varepsilon, 0)$-Stackelberg equilibrium:
\begin{align}
    \val[\outerset](\avgouter[][\numiters]) \leq \obj(\avgouter[][\numiters], \inner^*(\avgouter[][\numiters])) &\leq \min_{\outer \in \outerset} \val[\outerset](\outer) + \varepsilon\\
    \max_{\inner \in \innerset: \constr(\avgouter[][\numiters], \inner)} \obj(\avgouter[][\numiters], \inner) \leq \obj(\avgouter[][\numiters], \inner^*(\avgouter[][\numiters])) &\leq \min_{\outer \in \outerset} \max_{\inner \in \innerset : \constr(\outer, \inner)} \obj(\outer, \inner) + \varepsilon
\end{align}
\end{proof}

\begin{proof}[Proof of \Cref{thm:stackelberg-equiv}]
\sdeni{}{We can relax the inner player's payoff maximization problem via the problem's Lagrangian and since by \cref{main-assum}, Slater's condition is satisfied, strong duality holds, giving us for all $\outer \in \outerset$: \\ $\max_{\inner \in \innerset : \constr(\outer, \inner) \geq \zeros} \obj(\outer, \inner)  = \max_{\inner \in \innerset }  \min_{\langmult \geq \zeros} \lang[\outer]( \inner, \langmult) \\ =
\min_{\langmult \geq \zeros} \max_{\inner \in \innerset }  \lang[\outer]( \inner, \langmult)$.
We can then re-express the min-max game as: $\min_{\outer \in \outerset} \max_{\inner \in \innerset : \constr(\outer, \inner) \geq \zeros} \obj(\outer, \inner) = \min_{\langmult \geq \zeros} \min_{\outer \in \outerset} \max_{\inner \in \innerset } \\ \lang[\outer]( \inner, \langmult)$. Letting $\langmult^* \in \argmin_{\langmult \geq \zeros} \min_{\outer \in \outerset} \max_{\inner \in \innerset } \lang[\outer]( \inner, \langmult)$, we have $\min_{\outer \in \outerset} \\ \max_{\inner \in \innerset : \constr(\outer, \inner) \geq \zeros} \obj(\outer, \inner) = \min_{\outer \in \outerset} \max_{\inner \in \innerset }  \lang[\outer]( \inner, \langmult^*)$. Note that $\lang[\outer]( \inner, \langmult^*)$ is convex-concave in $(\outer, \inner)$. Hence, any Stackelberg equilibrium $(\outer^*, \inner^*) \in \outerset \times \innerset$ of $(\outerset, \innerset, \obj, \constr)$ is a saddle point of $\lang[\outer]( \inner, \langmult^*)$, i.e., $\forall \outer \in \outerset, \inner \in \innerset, \lang[\outer^*]( \inner, \langmult^*) \leq \lang[\outer^*]( \inner^*, \langmult^*) \leq \lang[\outer]( \inner^*, \langmult^*)$.}
\end{proof}

\begin{proof}[Proof of \Cref{thm:lang-regret-bound}]
Since the Lagrangian regret is bounded for both players we have:
\begin{align}
    &\left\{
    \begin{array}{c}
        \max_{\outer \in \outerset} \langregret[\outerset][\numiters](\outer) \leq \varepsilon\\
        \max_{\inner \in \innerset} \langregret[\innerset][\numiters](\inner) \leq \varepsilon
    \end{array}\right.\\
    &\left\{
    \begin{array}{c}
         \frac{1}{\numiters}\sum_{\iter = 1}^\numiters \lang[{\outer[ ][\iter]}][\iter](\inner[][\iter], \langmult^*) - \min_{\outer \in \outerset} \frac{1}{\numiters} \sum_{\iter =1}^\numiters  \lang[\outer][\iter] (\inner[][\iter],\langmult^*) \leq \varepsilon\\
        \max_{\inner \in \innerset} \frac{1}{\numiters} \sum_{\iter =1}^\numiters \lang[{\outer[][\iter]}][\iter](\inner, \langmult^*) -  \frac{1}{\numiters}\sum_{\iter = 1}^\numiters \lang[{\outer[][\iter]}][\iter](\inner[][\iter], \langmult^*) \leq \varepsilon
    \end{array}\right.\\
    &\left\{
    \begin{array}{c}
         \frac{1}{\numiters}\sum_{\iter = 1}^\numiters \lang[{\outer[ ][\iter]}](\inner[][\iter], \langmult^*) - \min_{\outer \in \outerset} \frac{1}{\numiters} \sum_{\iter =1}^\numiters  \lang[\outer] (\inner[][\iter],\langmult^*) \leq \varepsilon\\
        \max_{\inner \in \innerset} \frac{1}{\numiters} \sum_{\iter =1}^\numiters \lang[{\outer[][\iter]}](\inner, \langmult^*) -  \frac{1}{\numiters}\sum_{\iter = 1}^\numiters \lang[{\outer[][\iter]}](\inner[][\iter], \langmult^*) \leq \varepsilon
    \end{array}\right.
\end{align}

\noindent
The last line follows because the min-max Stackelberg game is static.

Summing the final two inequalities yields:
\begin{align}
    \max_{\inner \in \innerset} \frac{1}{\numiters} \sum_{\iter =1}^\numiters \lang[{\outer[][\iter]}] (\inner, \langmult^*) - \min_{\outer \in \outerset} \frac{1}{\numiters} \sum_{\iter=1}^\numiters  \lang[\outer] (\inner[][\iter], \langmult^*) \leq 2\varepsilon \\
    \frac{1}{\numiters} \sum_{\iter =1}^\numiters \max_{\inner \in \innerset} \lang[{\outer[][\iter]}] (\inner, \langmult^*) - \frac{1}{\numiters} \sum_{\iter=1}^\numiters \min_{\outer \in \outerset} \lang[\outer] (\inner[][\iter], \langmult^*) \leq 2\varepsilon
\end{align}
\noindent
where the second inequality was obtained by an application of Jensen's inequality on the first and second terms.

Since $\lang$ is convex in $\outer$ and concave in $\inner$, we have that $\max_{\inner \in \innerset}\\ \lang[{\outer[][\iter]}](\inner, \langmult^*)$ is convex in $\outer$ and $\min_{\outer \in \outerset}  \lang[\outer] (\inner[][\iter],\langmult^*)$ is convex in $\inner$, which implies that
    $\max_{\inner \in \innerset}  \lang[{\avgouter[][\numiters]}](\inner, \langmult^*) - \min_{\outer \in \outerset}  \lang[\outer] (\avginner[][\numiters],\langmult^*) \leq 2\varepsilon$.
By the max-min inequality (\cite{boyd2004convex}, Equation 5.46), it also holds that 
$\min_{\outer \in \outerset} \lang[\outer] (\avginner[][\numiters],\langmult^*) \leq \max_{\inner \in \innerset}  \lang[{\avgouter[][\numiters]}](\inner, \langmult^*)$.
Combining these two inequality yields the desired result.
\end{proof}

\begin{proof}[Proof of \Cref{thm:robustness_gd}]
The value function of the outer player in the game $\left\{(\outerset, \innerset, \obj[\iter]) \right\}_{\iter = 1}^\numiters$ at iteration $\iter \in \iters$,  is given by $\val[][\iter](\outer) = \max_{\inner \in \innerset} \obj[\iter](\outer, \inner)$. Hence, for all $\iter \in \iters$, as $\obj[\iter]$ is $\mu$-strongly-convex, $\val[][\iter]$ is also strongly concave since the maximum preserves strong-convexity.

Additionally, since for all $\iter \in \iters$, $\obj[\iter]$ is strictly concave in $\inner$, by Danskin's theorem \cite{danskin1966thm}, for all $\iter \in \iters$, $\val[][\iter]$ is differentiable and its derivative is given by $\grad[\outer] \val[][\iter](\outer) = \grad[\outer] \obj(\outer, \inner^*(\outer))$ where $\inner^*(\outer) \in \max_{\inner \in \innerset} \obj[\iter](\outer, \inner)$. Thus, as $\grad[\outer] \obj(\outer, \inner^*(\outer))$ is $\lipschitz[{\grad\obj}]$-lipschitz continuous, so is $\grad[\outer] \val[][\iter](\outer)$. The result follows from \citeauthor{cheung2019tracing}'s bound for gradient descent on shifting strongly convex functions (\cite{cheung2019tracing}, Proposition 12).

\end{proof}

\begin{proof}[Proof of \Cref{thm:robustness_lgda}]
By the assumptions of the theorem, the loss functions of the outer player $\{ \obj[\iter](\cdot, \inner[][\iter])\}_{\iter =1}^\numiters$ are $\mu_\outer$-strongly-convex and $\lipschitz[{\grad \obj}]$-Lipschitz continuous functions. Similarly the loss functions of the inner player $\{ - \obj[\iter](\outer[][\iter], \cdot)\}_{\iter =1}^\numiters$ are $\mu_\inner$-strongly-convex and $\lipschitz[{\grad \obj}]$-Lipschitz continuous functions. Using \citeauthor{cheung2019tracing}'s Proposition 12 \cite{cheung2019tracing}, we then obtain the following bounds:
\begin{align}
\left\|{\outer[][\numiters]}^* - \outer[][\numiters]\right\|    \leq (1 - \delta_\outer)^{\nicefrac{\numiters}{2}} \left\|{\outer[][0]}^* - \outer[][0]\right\| 
+ \sum_{\iter = 1}^\numiters \left( 1 - \delta_\outer \right)^{\frac{\numiters - \iter}{2}} \Delta_\outer^{(\iter)} \\
 \left\|{\inner[][\numiters]}^* - \inner[][\numiters]\right\|    \leq   (1 - \delta_\inner)^{\nicefrac{\numiters}{2}} \left\|{\inner[][0]}^* - \inner[][0]\right\| 
+ \sum_{\iter = 1}^\numiters \left( 1 - \delta_\inner \right)^{\frac{\numiters - \iter}{2}} \Delta_\inner^{(\iter)}
\end{align}

Combining the two inequalities, we obtain:

\begin{align}
&\left\|{\outer[][\numiters]}^* - \outer[][\numiters]\right\| + \left\|{\inner[][\numiters]}^* - \inner[][\numiters]\right\|    \notag \\
&\leq (1 - \delta_\outer)^{\nicefrac{\numiters}{2}} \left\|{\outer[][0]}^* - \outer[][0]\right\| + (1 - \delta_\inner)^{\nicefrac{\numiters}{2}} \left\|{\inner[][0]}^* - \inner[][0]\right\| \notag \\ 
&+ \sum_{\iter = 1}^\numiters \left( 1 - \delta_\outer \right)^{\frac{\numiters - \iter}{2}} \Delta_\outer^{(\iter)} + \sum_{\iter = 1}^\numiters \left( 1 - \delta_\inner \right)^{\frac{\numiters - \iter}{2}} \Delta_\inner^{(\iter)}
\end{align}

The second part of the theorem follows by taking the sum of the geometric series.
\end{proof}
\newpage
\section{Pseudo-Code for Algorithms}\label{sec-app:algos}

\begin{algorithm}[H]
\caption{Max-Oracle Gradient Descent}
\label{alg:mogd}
\textbf{Inputs:} $\outerset, \innerset, \obj, \constr, \learnrate, \outeriters, \outer^{(0)}$ \\ 
\textbf{Output:} $\outer^{*}, \inner^{*}$
\begin{algorithmic}[1]
\For{$\outeriter = 1, \hdots, \outeriters$}
    \State Find $\inner^*(\outer[][\iter -1]) \in \br[\innerset](\outer[][\iter -1])$ 
    \State Set $\inner^{(\outeriter-1)} = \inner^*(\outer[][\iter -1])$ 
    \State Set $\langmult^{(\outeriter-1)} = \langmult^*(\outer^{(\outeriter-1)}, \inner^{(\outeriter-1)})$
    \State Set $\outer^{(\outeriter)} = \project[\outerset] \left[ \outer^{(\outeriter-1)} - \learnrate[\outeriter] \grad[\outer] \lang[{\outer^{(\outeriter-1)}}]\left( \inner^{(\outeriter-1)}, \langmult^{(\outeriter-1)}\right) \right]$
\EndFor
\State Set $\avgouter[][\numiters] = \frac{1}{\numiters} \sum_{\iter = 1}^\numiters \outer[][\iter]$
\State Set $\inner^*(\avgouter[][\numiters]) \in \br[\innerset](\avgouter[][\numiters])$
\State \Return $(\avgouter[][\numiters], \inner^*(\avgouter[][\numiters]))$
\end{algorithmic}
\end{algorithm}

\begin{algorithm}[H]
\caption{Lagrangian Gradient Descent Ascent (LGDA)}
\label{alg:lgda}
\textbf{Inputs:} $\langmult^*, \outerset, \innerset, \obj, \constr,  \learnrate[][\outer], \learnrate[][\inner], \numiters,  \outer^{(0)}, \inner^{(0)}$ \\ 
\textbf{Output:} $\outer^{*}, \inner^{*}$
\begin{algorithmic}[1]
\For{$\iter = 1, \hdots, \numiters -1$}    
    \State Set $\outer^{(\iter +1)} = \project[\outerset] \left( \outer^{(\iter)} - \learnrate[\iter][\outer] \grad[\outer] \lang[{\outer[][\iter]}](\inner[][\iter], \langmult^*)
    \right)$

    \State Set $\inner^{(\iter +1)} = \project[{
    \innerset
    }] \left( \inner^{(\iter)} + \learnrate[\iter][\inner] \grad[\inner] \lang[{\outer[][\iter]}](\inner[][\iter], \langmult^*)
    \right)$ 
\EndFor
\State \Return $\{(\outer[][\iter], \inner[][\iter])\}_{\iter= 1}^\numiters$
\end{algorithmic}
\end{algorithm}

\begin{algorithm}[H]
\caption{Dynamic t\^atonnement}
\label{alg:dynamic_max_oracle_gd}
\textbf{Inputs:} $\numiters, \{(\util^{(\iter)}, \budget^{(\iter)}, \supply^{(\iter)}) \}_{\iter =1}^\numiters, \learnrate,  \price^{(0)}, \delta$ \\ 
\textbf{Output:} $\outer^{*}, \inner^{*}$
\begin{algorithmic}[1]
\For{$\iter = 1, \hdots, \numiters -1$}
    \State For all $\buyer \in \buyers$, find $\allocation[\buyer]^{(t)} \in \argmax_{\allocation[\buyer] \in \R^\numgoods_+:\allocation[\buyer]\cdot \price^{(\iter-1)} \leq \budget[\buyer]^{(\iter)}} \util[\buyer](\allocation[\buyer])$
    \State Set $\price^{(\iter)} =\project[\R_+^\numgoods]\left( \price^{(t-1)} - \learnrate[t](\supply^{(\iter)} - \sum_{\buyer \in \buyers} \allocation[\buyer]^{(t)})
    \right)$
\EndFor
\State \Return $(\price^{(\iter)}, \allocation^{(\iter)})_{\iter = 1}^\numiters$ 
\end{algorithmic}
\end{algorithm}

\begin{algorithm}[H]
\caption{Dynamic Myopic Best-Response Dynamics}
\label{alg:dynamic_lgda}
\textbf{Inputs:} $\{(\util^{(\iter)}, \budget^{(\iter)}, \supply^{(\iter)}) \}_{\iter =1}^\numiters, \learnrate[][\price], \learnrate[][\allocation], \numiters, \allocation^{(0)}, \price^{(0)}$ \\ 
\textbf{Output:} $\outer^{*}, \inner^{*}$
\begin{algorithmic}[1]
\For{$\iter = 1, \hdots, \numiters -1$}
    \State Set $\price^{(\iter +1)} = \project[\R_+^\numgoods]\left(
    \price^{(t)} - \learnrate[t][\price](\supply^{(\iter)} - \sum_{\buyer \in \buyers} \allocation[\buyer]^{(t)})
    \right)$
    \State For all $\buyer \in \buyers$, set $\allocation[\buyer]^{(\iter +1)} = \project[\R^\numgoods_+] \left( \allocation[\buyer]^{(\iter)} + \learnrate[\iter][\allocation]  \left( \frac{\budget[\buyer]^{(\iter)}}{\util[\buyer]^{(\iter)}\left(\allocation[\buyer]^{(\iter)}\right)} \grad[{\allocation[\buyer]}] \util[\buyer]^{(\iter)}\left(\allocation[\buyer]^{(\iter)}\right) - \price^{(\iter)} \right)\right)$
\EndFor
\State \Return $(\price^{(\iter)}, \allocation^{(\iter)})_{\iter = 1}^\numiters$ 
\end{algorithmic}
\end{algorithm}
\newpage
\section{An Economic Application: Details}\label{sec-app:fisher}

Our experimental goal was to understand if \Cref{alg:dynamic_max_oracle_gd}  and \Cref{alg:dynamic_lgda} converges in terms of distance to equilibrium and if so how the rate of convergences changes under different utility structures, i.e. different smoothness and convexity properties of the value functions. 

To answer these questions, we ran multiple experiments, each time recording the prices and allocations computed by \Cref{alg:dynamic_max_oracle_gd}, in the asymmetric learning setting, and by \Cref{alg:dynamic_lgda}, in the {symmetric} learning setting, during each iteration $t$ of the loop. Moreover, at each iteration $t$, we solve the competitive equilibrium $(\price^{(\iter)^\star}, \allocation^{(\iter)^\star})$ for the Fisher market $(\util^{(\iter)}, \budget^{(\iter)}, \supply^{(\iter)})$. 
Finally, for each run of the algorithm on each market, we then computed distance between the computed prices, allocations and the equilibrium prices, allocations, which we plot in \Cref{fig:exp_results_gd} and \Cref{fig:exp_results_lgda}. 

\paragraph{Hyperparameters}
We set up 100 different linear, Cobb-Douglas, Leontief {online} Fisher markets with random changing market parameters across time, each with $5$ buyers and $8$ goods, and we randomly pick one of these experiments to graph.  

In our execution of \Cref{alg:dynamic_max_oracle_gd}, 
buyer $\buyer$'s budget at iteration $t$, $\budget[\buyer]^{(\iter)}$, was drawn randomly from a uniform distribution ranging from $10$ to $20$ (i.e., $U[10,20]$),  each buyer $\buyer$'s valuation for good $\good$ at iteration $t$, $\valuation[i][j]^{(\iter)}$, was drawn randomly from $U[5,15]$, while each good $\good$'s supply at iteration $t$, $\supply[\good]^{(\iter)}$, was drawn randomly from $U[100,110]$.
In our execution of \Cref{alg:dynamic_lgda}, 
buyer $\buyer$'s budget at iteration $t$, $\budget[\buyer]^{(\iter)}$, was drawn randomly from a uniform distribution ranging from $10$ to $15$ (i.e., $U[10,15]$),  each buyer $\buyer$'s valuation for good $\good$ at iteration $t$, $\valuation[i][j]^{(\iter)}$, was drawn randomly from $U[10,20]$, while each good $\good$'s supply at iteration $t$, $\supply[\good]^{(\iter)}$, was drawn randomly from $U[10,15]$.

We ran both \Cref{alg:dynamic_max_oracle_gd} and \Cref{alg:dynamic_lgda} for 1000 iterations 
on linear, Cobb-Douglas, and Leontief Fisher markets.
We started the algorithm with initial prices drawn randomly from $U[5,55]$.
%
After manual hyper-parameter tuning, for \Cref{alg:dynamic_max_oracle_gd}, we opted for $\forall \iter \in \iters, \learnrate[\iter] = \frac{1}{\sqrt{t}}$ for all of linear, Cobb-Douglas, and Leontief Fisher markets. Moreover, for \Cref{alg:dynamic_lgda}, we opted for a {online} learning rate of $\forall \iter \in \iters, \learnrate[\iter][\outer] = \frac{5}{\sqrt{t}}$, $\learnrate[\iter][\inner] = \frac{0.01}{\sqrt{t}}$ for all of Linear, Cobb-Douglas, and Leontief Fisher markets.

\paragraph{Programming Languages, Packages, and Licensing}
We ran our experiments in Python 3.7 \cite{van1995python}, using NumPy \cite{numpy}, Pandas \cite{pandas}, and CVXPY \cite{diamond2016cvxpy}.
\Cref{fig:exp_results_gd} and \Cref{fig:exp_results_lgda} were graphed using Matplotlib \cite{matplotlib}.

Python software and documentation are licensed under the PSF License Agreement. Numpy is distributed under a liberal BSD license. Pandas is distributed under a new BSD license. Matplotlib only uses BSD compatible code, and its license is based on the PSF license. CVXPY is licensed under an APACHE license. 

\paragraph{Implementation Details}
In order to project each allocation computed onto the budget set of the consumers, i.e., $\{\allocation \in \R^{\numbuyers \times \numgoods}_+ \mid \allocation\price \leq \budget\}$, we used the alternating projection algorithm for convex sets, and alternatively projected onto the sets $\R^{\numbuyers \times \numgoods}_+$ and $\{\allocation \in \R^{\numbuyers \times \numgoods} \mid \allocation\price \leq \budget\}$. 

To compute the best-response for the inner play in \Cref{alg:dynamic_max_oracle_gd}, we used the ECOS solver, a CVXPY’s first-order convex-program solvers, but if ever a runtime exception occurred, we ran the SCS solver.

When computing the distance from the demands $\allocation^{(\iter)}$ computed by our algorithms to the equilibrium demands $\allocation^{(\iter)^\star}$, we normalize both demands to satisfy $\forall \good \in \goods, \;\sum_{\buyer \in \buyers} \allocation[i][j]=1_{\numgoods}$ to reduce the noise caused by changing supplies.

\paragraph{Computational Resources}
Our experiments were run on MacOS machine with 8GB RAM and an Apple M1 chip, and took about 2 hours to run. Only CPU resources were used.

\paragraph{Code Repository}
The data our experiments generated, and the code used to produce our visualizations, can be found in our code repository ({\color{blue}\rawcoderepo}).

\end{document}